\documentclass[aps,prb, twocolumn, superscriptaddress, amsmath, amsfonts, amssymb, amsthm]{revtex4-2}
\usepackage{graphicx}
\usepackage{dcolumn}
\usepackage[unicode=true,
 bookmarks=false,
 breaklinks=true,pdfborder={0 0 1},backref=false,colorlinks=true]
 {hyperref}
\hypersetup{pdfcreator={},pdfproducer={LaTeX with hyperref},linkcolor=blue,anchorcolor=blue,citecolor=blue,filecolor=red,menucolor=red,urlcolor=blue,pdfstartview=FitV,pdfhighlight=/I,hypertexnames=true}
\makeatletter
\usepackage{lmodern}
\usepackage{mathptmx, newtxtext, newtxmath, xspace}
\usepackage{colortbl}
\usepackage{array}
\usepackage{bbm}
\usepackage{siunitx}
\usepackage{ulem}
\usepackage{comment}
\usepackage[table, dvipsnames]{xcolor}

\usepackage{bm}
\usepackage{braket}

\definecolor{colorA}{cmyk}{0,0,0,0.05}
\definecolor{colorB}{cmyk}{0.14,0.04,0,0}
\definecolor{colorC}{cmyk}{0.02,0.0799,0,0}
\definecolor{colorD}{cmyk}{0.099,0.14,0,0}
\definecolor{FMirrepColour}{gray}{0.4}

\usepackage{natbib}
\setcitestyle{numbers}

\makeatother

\begin{document}
\title{Holographic superconductivity of a critical Fermi surface}
\author{Veronika C. Stangier }
\affiliation{Institute for Theory of Condensed Matter, Karlsruhe Institute of Technology,
Karlsruhe 76131, Germany}
\author{J\"org Schmalian }
\affiliation{Institute for Theory of Condensed Matter, Karlsruhe Institute of Technology,
Karlsruhe 76131, Germany}
\affiliation{Institute for Quantum Materials and Technologies, Karlsruhe Institute
of Technology, Karlsruhe 76131, Germany}
\date{\today }
\begin{abstract}
We derive a holographic formulation of triplet superconductivity in a two-dimensional metal at a ferromagnetic quantum critical point. Starting from a large-$N$ Yukawa-Sachdev-Ye-Kitaev  model of compressible fermions coupled to quantum-critical Ising ferromagnetic fluctuations, we reformulate the pairing problem in terms of bilocal collective fields and analyze Gaussian fluctuations around the quantum-critical normal state. We demonstrate that the resulting pairing action can be mapped onto a scalar field theory in an emergent curved spacetime with ${\rm AdS}_{2}\otimes\mathbb{R}_{2}$ geometry. The additional holographic dimension is shown to encode the internal dynamics of Cooper pairs and is related nonlocally to the frequency dependence of the anomalous Gor’kov function via a Radon transform. Within this framework, the onset of superconductivity corresponds to a Breitenlohner–Freedman instability of the scalar field, which is shown to be equivalent to the pairing instability obtained from the linearized Eliashberg equations. The factorized ${\rm AdS}_{2}\otimes\mathbb{R}_{2}$ geometry reflects the local-in-space but critical-in-time character of fermionic excitations near a metallic quantum critical point and corresponds to what one expects in the vicinity of a Reissner-Nordstr\"om black hole. Our results provide a microscopic derivation of holographic superconductivity in a compressible quantum critical metal and clarify the geometric structure underlying quantum-critical pairing.

\end{abstract}
\maketitle
\section{Introduction}
 Superconductivity in the vicinity of a quantum critical point~\cite{Sachdevbook} (QCP) differs qualitatively from its behavior in a well-established Fermi liquid, where the Cooper instability~\cite{Cooper1956,BCS1957a,BCS1957b} and the Kohn-Luttinger mechanism~\cite{Kohn1965} render the superconducting state the natural ground state of a metal. Near a QCP, strong fluctuations tend to destroy well-defined quasiparticles while simultaneously generating singular interactions. The interplay of these two effects - ill-defined fermionic constituents that are nevertheless expected to form Cooper pairs, subject to a strongly retarded and resonant interaction - lies at the heart of what is commonly referred to as quantum-critical pairing.

Quantum-critical pairing has been studied  within suitable generalizations of Eliashberg theory~\cite{Eliashberg1960} to fermions interacting with critical bosonic modes~\cite{Bonesteel1996,Son1999,Abanov2001,Abanov2001b,Roussev2001,Chubukov2003,Abanov2003,Chubukov2005,She2009,Moon2010,Levchenko2013,Wang2013,Wang2015,Varma2016,Khodas2020,Metlitski2015,Fitzpatrick2015,Raghu2015,Mandal2016,Wang2016,Wu2019,Esterlis2019,Wang2020,Abanov2020,Valentinis2023a,Valentinis2023b,Esterlis2026}. As a result, a strongly dynamical pairing state emerges that is stabilized in the strong-coupling regime and has been shown to generically yield higher transition temperatures at the QCP than away from criticality; see e.g. Refs.~\cite{Valentinis2023a,Valentinis2023b}.  The appeal of this approach is that it allows one to start from a concrete microscopic model that incorporates both the electronic band structure and the dominant soft collective mode or gauge excitation. In this way, one obtains information about a well-defined microscopic Hamiltonian describing, for example, a ferromagnetic, spin- or charge-density-wave, or nematic system with a given band structure.

An alternative perspective on quantum-critical pairing is provided by the framework of holographic superconductivity~\cite{Gubser2008,Hartnoll2008,Hartnoll2008b}, which exploits the holographic correspondence between a $d+1$-dimensional quantum field theory and a gravity theory in $d+2$ dimensions,  with asymptotic anti-de Sitter (${\rm AdS}$) spacetime~\cite{Maldacena1997,Witten1998,Gubser1998}. Holography has emerged as a powerful tool for understanding strongly coupled many-body systems~\cite{Sachdev2012,Zaanen2015,Ammon2015,Hartnoll2018,Baggioli2019}. It relates theories of strongly interacting particles to gravitational theories in a higher-dimensional spacetime, where the additional holographic dimension is not manifest in the original field theory. Even in situations where quasiparticle descriptions break down, the correspondence enables controlled insights into transport phenomena~\cite{Kovtun2005,Hartnoll2007,Hartnoll2015,Blake2016} and provides a unified framework for describing broken-symmetry states such as charge-density waves~\cite{Donos2011,Delacretaz2017} or superconductivity~\cite{Gubser2008,Hartnoll2008,Hartnoll2008b}. The application of the approach to condensed-matter physics problems is largely phenomenological, relying primarily on symmetry considerations and being comparatively insensitive  to microscopic details of the critical metal.

Despite the remarkable power of holographic duality, it remains important to clarify the microscopic origin of gravitational formulations of quantum many-body problems and to elucidate the physical meaning of the extra holographic dimension in specific condensed-matter systems. In particular, establishing a  connection to quantum-critical Eliashberg theory could be beneficial for both approaches to quantum-critical pairing. One promising route toward addressing these questions is to derive holography from concrete many-body models. Such a program would also enable a more explicit application of holographic methods to strong-coupling problems. Progress in this direction has been achieved for the zero-dimensional Sachdev-Ye-Kitaev (SYK) model~\cite{Sachdev1993,Georges2001,Kitaev2015,Kitaev2015b}. At low energies, the $0+1$-dimensional SYK model is governed by the same effective theory as two-dimensional gravity of anti-de Sitter space ${\rm AdS}_2$ without matter fields~\cite{Sachdev2015,Maldacena2016,Das2018}. Another advance is Ref.~\cite{Inkof2022}, where holographic superconductivity in ${\rm AdS}_{2}$ was derived from the superconducting Yukawa-SYK model of Ref.~\cite{Esterlis2019}. As anticipated phenomenologically within the framework of holographic superconductivity~\cite{Gubser2008,Hartnoll2008,Hartnoll2008b}, a scalar field emerges as a matter degree of freedom in ${\rm AdS}_{2}$. Ref.~\cite{Inkof2022} established a direct correspondence between this scalar field and the Gork'ov function~\cite{Gorkov1958}, which is widely used in the field-theoretical description of superconductors~\cite{Abrikosow1964,Mahan2013}. At finite temperatures, the emergent spacetime obtained from the many-body analysis contains a black hole with the associated Bekenstein temperature. In particular, a close correspondence between the solution of quantum-critical pairing within Eliashberg theory and holographic pairing in ${\rm AdS}_{2}$ was demonstrated~\cite{Inkof2022}; see also Ref.~\cite{Esterlis2026} for a recent review.

Extending such derivations to systems with finite spatial dimensionality would reveal whether and how quantum-critical temporal and spatial scales become intertwined. In this context, one may distinguish between two qualitatively different classes of systems: critical metals with a Fermi surface, corresponding to compressible states of matter, and systems without a Fermi surface. Recent work~\cite{Stangier2026,Stangier2026c} has shown that quantum-critical pairing can also occur in Dirac systems at zero density near the Gross–Neveu transition~\cite{Gross1974,ZinnJustin1991}. Comparing the emergence of gravitational descriptions in these distinct settings is expected to shed further light on their differing physical behavior.

In this paper, we derive a holographic description of superconductivity at, or in the vicinity of, a ferromagnetic quantum critical point. The system is described by a two-dimensional model of compressible electrons with a filled Fermi sea interacting with a soft Ising-ferromagnetic collective mode. We obtain a gravitational formulation in terms of a spacetime of the form ${\rm AdS_{2}\otimes\mathbb{R}_{2}}$, where the temporal dynamics is governed by a nontrivial geometry while the spatial sector remains flat. This result is consistent with expectations based on the Reissner–Nordstr\"om black-hole geometry for systems with fixed charge density in holographic formulations~\cite{Sachdev2012,Zaanen2015,Ammon2015,Hartnoll2018,Baggioli2019,Iqbal2012,Kachru2008}. Specifically, we find that superconductivity in the critical state is described by a Ginzburg–Landau theory
\begin{equation}
S_{\rm sc}=\int d^{4}\xi\sqrt{g}\left(\partial_{\mu}\psi^{}\partial^{\mu}\psi+m^2\psi^{}\psi\right)\label{eq:HS_GL}
\end{equation}
defined in an emergent four-dimensional spacetime with coordinates $\xi^{\mu}=(\boldsymbol{x},\tau,\zeta)$ and metric
\begin{equation}
ds^{2}=g_{\mu\nu}d\xi^{\mu}d\xi^{\nu}=\frac{d\zeta^{2}+d\tau^{2}}{\zeta^{2}}+k_{\rm F}^2d\boldsymbol{x}^{2}.\label{eq:metric}
\end{equation}
Here $m$ denotes the mass of the collective field, $\boldsymbol{x}$ represents the two-dimensional spatial coordinates, and $\tau$ and $\zeta$ are related to the center-of-mass and relative imaginary times of fluctuating Cooper pairs within the Matsubara formalism. The metric in Eq.~\eqref{eq:metric} corresponds to a Eucledian ${\rm AdS_{2}}\otimes\mathbb{R}_{2}$ spacetime, while $k_{\rm F}$ is the Fermi momentum. The extra holographic coordinate $\zeta$ describes, in close analogy to the zero-dimensional case discussed in Ref.~\cite{Inkof2022}, the internal dynamics of Cooper pairs formed out of the quantum-critical normal state. In Ref.~\cite{Stangier2026b}, a related holographic theory for two-dimensional Dirac fermions at the Gross–Neveu transition will be presented, demonstrating that this problem can be mapped onto a holographic superconductor in ${\rm AdS}_4$.

The Ginzburg–Landau theory in Eq.~\eqref{eq:HS_GL} becomes unstable toward condensation, signaling the onset of superconductivity, once $m^2$ drops below a critical threshold. In flat space this instability occurs for $m^2=0$. However, as shown by Breitenlohner and Freedman~\cite{Breitenlohner1982}, in negatively curved spacetime condensation sets in only when $m^2=m^2_{\rm BF}<0$. In ${\rm AdS}_2$, the Breitenlohner–Freedman bound is $m^2_{\rm BF}=-1/4$. Below, we demonstrate that this instability coincides with the onset of pairing obtained from the instability of the linearized Eliashberg equations. Furthermore, we obtain explicit expressions for the mass $m$ in Eq.~\eqref{eq:HS_GL}, determined by microscopic parameters of the microscopic model of an Ising ferromagnet.

In Secs.~\ref{sec:model} and~\ref{sec:Stationary Solution} we formulate and solve the many-body problem in a suitable large-$N$ limit. Within this framework we reproduce known results for the quantum-critical normal state and for superconductivity, employing a formulation in terms of bilocal collective fields. In Sec.~\ref{sec:Pairing Fluctuations} we analyze Gaussian pairing fluctuations around the saddle point. Finally, in Sec.~\ref{sec:Holographic Map} we construct the explicit holographic mapping. This mapping, presented in Eq.~\eqref{eq:the_map}, constitutes the central result of the present work. Our findings show that, for metallic quantum-critical states, the holographic correspondence established for the SYK model~\cite{Inkof2022} can be extended to more realistic systems in finite spatial dimensions, thereby establishing direct contact with established results in condensed-matter physics.

\section{The model}
\label{sec:model}

The Hamiltonian of our analysis describes a two-dimensional ($d=2$) system of fermions coupled to quantum-critical Ising-ferromagnetic bosons, as discussed in Refs.~\cite{Chubukov2004,Rech2006,Damia2019,Roussev2001,Chubukov2003,Xu2017,Xu2020}:
\begin{eqnarray}
H & = & \sum_{\boldsymbol{p}i\sigma}\varepsilon_{\boldsymbol{p}}c_{\boldsymbol{p}i\sigma}^{\dagger}c_{\boldsymbol{p}i\sigma}+\frac{1}{2}\sum_{\boldsymbol{q}l}\left(\pi_{\boldsymbol{q}l}\pi_{-\boldsymbol{q}l}+\omega_{\boldsymbol{q}}^{2}\phi_{\boldsymbol{q}l}\phi_{-\boldsymbol{q}l}\right)\nonumber \\
 & + & \frac{1}{N}\sum_{\boldsymbol{p}\boldsymbol{q},ijl\sigma\sigma'}\check{g}_{ijl}c_{\boldsymbol{p}+\boldsymbol{q}i\sigma}^{\dagger}\sigma_{\sigma\sigma'}^{z}c_{\boldsymbol{p}j\sigma'}\phi_{-\boldsymbol{q}l}+h.c..\label{eq:Hamiltonian}
\end{eqnarray}
Here, $c_{\boldsymbol{p}i\sigma}^{\dagger}$ creates a fermion with momentum $\boldsymbol{p}$ and spin $\sigma$.
The fermionic dispersion is given by $\varepsilon_{\boldsymbol{p}}$, which we assume to be non-nested.
For each $\left(\boldsymbol{p},\sigma\right)$ there is an additional flavor index $i=1\cdots N$, introduced to enable a controlled large-$N$ formulation.
The field $\phi_{\boldsymbol{q}l}$ denotes a charge-neutral boson that is odd under time reversal, with momentum $\boldsymbol{q}$ and flavor index $l=1,\cdots,M$. Its bare dispersion is
\begin{equation}
\omega_{\boldsymbol{q}}^{2}\approx\omega_{0}^{2}+c^{2}q^{2}.
\end{equation}
The operator $\pi_{\boldsymbol{q}l}$ is the momentum conjugate to $\phi_{\boldsymbol{q}l}$.
The Yukawa coupling $\check{g}_{ijl}$ represents a random all-to-all interaction in flavor space. This follows the large-$N$, $M$ formulation of the zero-dimensional Yukawa-SYK model introduced in Ref.~\cite{Esterlis2019,Wang2020}. In the lattice realization, the couplings $\check{g}_{ijl}$ are identical on all lattice sites and at all times; hence, space- and time-translation invariance remain intact even for a given realization of $\check{g}_{ijl}$~\cite{Chowdhury2018,Esterlis2021,Kim2021}.
Models similar to Eq.~\eqref{eq:Hamiltonian} have been proposed for time-reversal-even bosons describing Ising-nematic states~\cite{Esterlis2021,Patel2022,Li2024,Guo2024}.
Corresponding multi-flavor problems involving Yukawa-coupled Dirac fermions were analyzed in Refs.~\cite{Kim2021,Stangier2026}, while related models with direct electron–electron interactions were studied in Ref.~\cite{Chowdhury2018}.
In Refs.~\cite{Kim2021,Patel2022} it was shown that, in the limit of large $N$ and $M$, these models exhibit maximally chaotic behavior.

We focus on the Ising-ferromagnetic case because it describes systems with spin-orbit coupling that are relevant to many correlated materials. Moreover, this choice avoids the superconducting first-order transition encountered in the Heisenberg limit~\cite{Chubukov2003}.
Fluctuations associated with the closely related coupling to nematic excitations, which would favor $s$-wave pairing, are suppressed by coupling to acoustic phonons~\cite{Karahasanovic2016,Paul2017}; this suppression is absent in the Ising-ferromagnetic problem. However, our analysis can be extended to pairing mediated by fluctuating altermagnetic excitations with  singlet pairing~\cite{Schmalian1998,Wu2025}.

The random coupling constants $\check{g}_{ijl}$ are taken to be complex Gaussian variables, with variances $\left(1-\frac{\alpha}{2}\right)\check{g}^{2}/2$ and $\alpha \check{g}^{2}/4$ for their real and imaginary parts, respectively.
We will show that sufficiently small $\alpha$ allows for a superconducting solution, whereas superconductivity is absent at large $N$ for $\alpha=1$. Thus, $\alpha$ acts as a pair-breaking parameter and determines the effective coupling strength in the pairing channel~\cite{Hauck2020}:
\begin{equation}
\check{g}_{p}^{2}=\check{g}^{2}\left(1-\alpha\right).
\label{eq:g_pair}
\end{equation}
We consider the limit of large $N$ and $M$, keeping the ratio $\mu=M/N$ finite.
The bare magnetic correlation length $\xi_{0}=c/\omega_{0}$ is renormalized by coupling to the fermions and diverges at a ferromagnetic quantum-critical point (QCP).
This occurs at a specific value of the coupling constant $g=g_c\sim \sqrt{\rho_{\rm F}} \omega_0$, placing the QCP in the strong-coupling regime of the model.
In the absence of randomness in $\check{g}_{ijl}$, Eq.~\eqref{eq:Hamiltonian} was analyzed in Refs.~\cite{Chubukov2004,Rech2006,Damia2019,Roussev2001,Chubukov2003,Xu2017,Xu2020}.
However, as shown in Refs.~\cite{Lee2008,Metlitski2010}, the large-$N$ formulation becomes technically involved in this regime, complications that are avoided within the approach adopted in the present work.

\subsection{Bilocal fields and large-$N$}
 The large-$N$ analysis of the
model follows closely Refs.~\cite{Esterlis2021,Chowdhury2018,Kim2021}.
Instead of formulating the problem in terms of the primary degrees of freedoms, i.e. the fermions and bosons of Eq.~\eqref{eq:Hamiltonian}, we introduce bilocal collective
fields
\begin{eqnarray}
G_{\sigma\sigma'}\left(x,x'\right) & = & \frac{1}{N}\sum_{i=1}^{N}c_{i\sigma}\left(x\right)c_{i\sigma'}^{\dagger}\left(x'\right),\nonumber \\
D\left(x,x'\right) & = & \frac{1}{M}\sum_{l=1}^{M}\phi_{l}\left(x\right)\phi_{l}\left(x'\right).\label{eq:bilocal}
\end{eqnarray}
$x=\left(\boldsymbol{x},\tau\right)$ comprises space and imaginary
time. For pair-breaking parameter $\alpha\neq1$ one must also include bilocal pairing fields
\begin{equation}
F_{\sigma\sigma'}\left(x,x'\right)=\frac{1}{N}\sum_{i=1}^{N}c_{i\sigma}\left(x\right)c_{i\sigma'}\left(x'\right)\label{eq:bilocalF}
\end{equation}
as well as $F^{\dagger}$ with $c$ replaced by $c^{\dagger}$~\cite{Esterlis2019}.
 The
usage of these bilocal fields will be very helpful for our formulation
of a holographic theory. Eqn.~\eqref{eq:bilocal} and
\eqref{eq:bilocalF} are enforced via Lagrange-multiplier fields $\Sigma$,
$\Pi$, and $\Phi$ that depend on the same set of coordinates. This allows
writing the averaged interaction term of the action as 
\begin{eqnarray*}
S_{{\rm int}} & = & \check{g}^{2}M\int_{x,x'}{\rm tr}\left(\hat{\sigma}^{z}\hat{G}\left(x,x'\right)\hat{\sigma}^{z}\hat{G}\left(x',x\right)\right)D\left(x,x'\right)\\
 & - & \check{g}_{p}^{2}M\int_{x,x'}{\rm tr}\left(\hat{\sigma}^{z}\hat{F}\left(x,x'\right)\hat{\sigma}^{z}\hat{F}^{\dagger}\left(x',x\right)\right)D\left(x,x'\right).
\end{eqnarray*}
${\rm tr}$ stands for the trace over spin indices, hats refer to
matrices in spin space, and $\int_{x}=\int d^{2}\boldsymbol{x}d\tau$ stands for the integration over the $2+1$ coordinates.
Now, the original fermions and bosons can be integrated out, yielding
a theory exclusively in terms of bilocal fields:
\begin{eqnarray}
S & = & -\frac{N}{2}{\rm \mathbf{Tr}}\log\left(\hat{\boldsymbol{g}}_{0}^{-1}-\hat{\boldsymbol{\Sigma}}\right)+\frac{M}{2}\text{Tr}\log\left(d_{0}^{-1}-\Pi\right)\nonumber \\
 & - & \frac{N}{2}{\rm \mathbf{Tr}}\,\hat{\boldsymbol{G}}\otimes\hat{\boldsymbol{\Sigma}}+\frac{M}{2}{\rm Tr}\hat{D}\otimes\hat{\Pi}+S_{{\rm int}}.\label{eq:act}
\end{eqnarray}

While similar to a Luttinger-Ward
functional~\cite{Luttinger1960,Abrikosow1964,Negele1988}, the bilocal
fields  are genuine dynamic variables of a collective field theory.
In Eq.\eqref{eq:act} we use ${\rm Tr}\hat{A}\otimes\hat{B}=\int_{xx'}{\rm tr}\left(\hat{A}\left(x,x'\right)\hat{B}\left(x',x\right)\right)$
and matrices in Nambu space
\begin{equation}
\hat{\boldsymbol{G}}\left(x,x'\right)=\left(\begin{array}{cc}
\hat{G}\left(x,x'\right) & \hat{F}\left(x,x'\right)\\
\hat{F}^{\dagger}\left(x,x'\right) & \tilde{G}\left(x,x'\right)
\end{array}\right),
\end{equation}
as well as
\begin{equation}
\hat{\boldsymbol{\Sigma}}\left(x,x'\right)=\left(\begin{array}{cc}
\hat{\Sigma}\left(x,x'\right) & \hat{\Phi}\left(x,x'\right)\\
\hat{\Phi}^{\dagger}\left(x,x'\right) & \tilde{\Sigma}\left(x,x'\right)
\end{array}\right).
\end{equation}
 ${\rm \mathbf{Tr}}$  in Eq.\eqref{eq:act}  stands for an additional
trace over the Nambu components. Finally, we used
with $\tilde{A}\left(x,x'\right)=-\hat{A}^{T}\left(x',x\right)$. $\hat{\boldsymbol{g}}_{0}$ and $d_{0}$ are the  bare fermion and boson propagators, respectively.

\section{Stationary Solution}
\label{sec:Stationary Solution}
At large-$N$, $M$
and fixed $\mu=M/N$ the saddle-point equations
\begin{equation}
\delta S/\delta G=0,\,\,\,\delta S/\delta \Sigma=0,\,\,\,\delta S/\delta F=0,\,\,\cdots
\end{equation}
become exact, allowing for an analysis at generic values of the coupling constant, including the strong-coupling limit.  At the saddle point, the fields only depend on
$x-x'$.
Fourier transformation
to momentum and frequency variables $p=\left(\boldsymbol{p},\epsilon\right)$, the saddle-point conditions $\delta S/\delta G=\delta S/\delta F^\dagger=\delta S/\delta D=0$
yield the coupled Eliashberg equations for the fermion self energy, supplemented by the bosonic self energy:
\begin{eqnarray}
\hat{\Sigma}\left(p\right) & = & -\mu \check{g}^{2}\int_{p'}\hat{\sigma}^{z}\hat{G}\left(p'\right)\hat{\sigma}^{z}D\left(p-p'\right),\nonumber \\
\hat{\Phi}\left(p\right) & = & \mu \check{g}_{p}^{2}\int_{p'}\hat{\sigma}^{z}\hat{F}\left(p'\right)\hat{\sigma}^{z}D\left(p-p'\right),\nonumber \\
\Pi\left(q\right) & = & -\check{g}^{2}\int_{p}{\rm tr}\left[\hat{\sigma}^{z}\hat{G}\left(p\right)\hat{\sigma}^{z}\hat{G}\left(p+q\right)\right]\nonumber \\
 & + & \check{g}_{p}^{2}\int_{p}{\rm tr}\left[\hat{\sigma}^{z}\hat{F}\left(p\right)\hat{\sigma}^{z}\hat{F}^{\dagger}\left(p+q\right)\right].\label{eq:Eliahb}
\end{eqnarray}
In addition, $\delta S/\delta \Sigma=\delta S/\delta \Phi^\dagger=\delta S/\delta \Pi=0$
yield the Dyson equations
 \begin{equation}
D\left(q\right)^{-1}=d_0\left(q\right)^{-1}-\Pi \left(q\right)
\end{equation}
 for the bosonic and 
\begin{equation}
\hat{\boldsymbol{G}}\left(p\right)^{-1}=\hat{\boldsymbol{g}}_{0}\left(p\right)^{-1}-\hat{\boldsymbol{\Sigma}}\left(p\right)
\end{equation}
for the fermionic propagators. Hence, at the saddle point the collective fields 
behave like propagators and self energies.
The large-$N$, $M$ limit corresponds to a self-consistent summation of one-loop diagrams.
These equations agree with, and might serve as a justification for the one-loop diagrammatic
treatments of Refs.~\cite{Chubukov2004,Rech2006,Damia2019,Roussev2001,Chubukov2003,Xu2017,Xu2020}.

In what follows we first discuss the saddle point solutions of the normal state at the QCP and then, in a second step, consider small, Gaussian fluctuations on top
of it, i.e. 
\begin{eqnarray}
    \hat{\boldsymbol{G}}&=&\hat{\boldsymbol{G}}_{{\rm sp}}+\delta\hat{\boldsymbol{G}} \nonumber \\
    D&=&D_{{\rm sp}}+\delta D, 
\end{eqnarray}
and similar for the conjugated self energies, where $\hat{\boldsymbol{G}}_{{\rm sp}}$ etc. are the solution of Eq.\eqref{eq:Eliahb}. We focus on Gaussian fluctuations
in $F$ and $\Phi$ which decouple from all other fluctuations due to
the $U\left(1\right)$ invariance of the normal state.

\subsection{Normal state saddle at the QCP} If the saddle point values
of $F$ and $\Phi$ vanish, the system is in its normal state. To
avoid cluttering equations we measure momenta in units of the Fermi
momentum $k_{\rm F}$ and energies in units of $\varepsilon_{F}=v_{F}k_{\rm F}$.
At the QCP, the bosonic and fermionic self energies take the form
\begin{eqnarray}
\Pi_{\boldsymbol{q}}\left(\epsilon\right) & = & \omega_{0}^{2}-2g^{2}\rho_{F}\frac{\left|\epsilon\right|}{\left| \boldsymbol{q}\right|},\label{eq:Pi_ns} \\
\Sigma_{\boldsymbol{p}}\left(\epsilon\right) & = & -i\lambda{\rm sign}\left(\epsilon\right)\left|\epsilon\right|^{2/3},\label{eq:Sigma_ns}
\end{eqnarray}
where $\rho_{F}$ is the density of states while 
\begin{equation}
    \lambda=\frac{\mu}{2\sqrt{3}}\left(\sqrt{2\rho_{F}}\check{g}\frac{v_{F}}{c}\right)^{4/3}
    \label{eq:lambda}
\end{equation}
is the dimensionless coupling constant. 

At low energy, the propagators are determined by $1/G_{\boldsymbol{p}}\left(\epsilon\right)\approx-\varepsilon_{\boldsymbol{p}}-\Sigma_{\boldsymbol{p}}\left(\epsilon\right)$
and $1/D_{\boldsymbol{q}}\left(\epsilon\right)\approx\omega_{\boldsymbol{q}}^{2}-\Pi_{\boldsymbol{q}}\left(\epsilon\right)$, i.e. the low-energy dynamics is determined by the frequency dependence of the self energies and not of the bare propagators.
The derivation of these results from Eq.~\eqref{eq:Eliahb} is well
established~\cite{Chubukov2004,Rech2006,Damia2019,Roussev2001,Chubukov2003}.
Eqs.~\eqref{eq:Pi_ns} and \eqref{eq:Sigma_ns} also agree  with findings
from sign-problem free Quantum Monte Carlo calculations for the corresponding
non-random Ising ferromagnet~\cite{Xu2017,Xu2020}. 


\subsection{Onset of superconductivity} 
The onset of superconductivity
at the QCP can be analyzed from the linearized gap equation for $\hat{\Phi}\left(p\right)=\sum_{j=0}^{3}\Phi_{\boldsymbol{p}}^{\left(j\right)}\left(\epsilon\right)i\hat{\sigma}^{y}\hat{\sigma}^{j}$.
One finds an attractive interaction of equal spin states, i.e. triplet
pairing with $j=1$ or $2$. For $\left|\left|\mathbf{p}\right|-k_{\rm F}\right|\ll k_{\rm F}$
the anomalous self energy only depends on the angle $\theta_{\boldsymbol{p}}$
of $\boldsymbol{p}$. For a rotationally invariant problem it can be expanded in harmonics, where $l\in\mathbb{Z}$
is the angular momentum, with gap equation diagonal in $l$. Then, different
angular momenta are almost degenerate, where the degeneracy is lifted
only due to effects that are sub-leading at low energies. Including
these sub-leading effects, the leading channel is $p$-wave triplet
pairing with $l=\pm$1. In lattice systems that break rotation invariance, the dominant pairing is in the irreducible representation of the point group that is odd under inversion and transforms like the in-plane coordinates. 
In  all these leading pairing  channels, the frequency dependence of
$\Phi\left(\epsilon\right)$ follows from the linearized Eliashberg
equation
\begin{equation}
\Phi\left(\epsilon\right)=\frac{1-\alpha}{3}\int d\epsilon'\frac{\Phi\left(\epsilon'\right)}{\left|\epsilon'\right|^{2/3}\left|\epsilon-\epsilon'\right|^{1/3}},\label{eq:gamma-1}
\end{equation}
and is solely determined by the pair-breaking parameter $\alpha$. An  equation  like this
was analyzed in Refs.~\cite{Son1999,Halboth2000,Abanov2001,Abanov2001b,Roussev2001,Chubukov2003,Abanov2003,Chubukov2005,She2009,Moon2010,Levchenko2013,Wang2013,Wang2015,Varma2016,Khodas2020,Metlitski2015,Fitzpatrick2015,Raghu2015,Mandal2016,Wang2016,Wu2019,Esterlis2019,Abanov2020,Esterlis2026}. Eq.~\eqref{eq:gamma-1} is obtained by integrating over momenta perpendicular to the Fermi surface and using that (i) fermions are slower than boson modes and (ii) the fermionic self energy is momentum independent. It is solved via the
powerlaw ansatz $\Phi(\epsilon )\sim | \epsilon |^{-1/6+i \beta}$ and the superconducting ground state survives  until $\beta \rightarrow 0$, i.e. until
$\alpha$ reaches a critical strength $\alpha^{*}$, determined by
\begin{equation}
    1=\tfrac{1}{3}\int_{-\infty}^{\infty}dx\frac{1-\alpha^{*}}{\left|1-x\right|^{1/3}\left|x\right|^{5/6}}.
    \label{eq:pairing_cond_gamma_1/3}
\end{equation}
This yields $\alpha_{*}\approx0.879618$. In the vicinity of this point the transition temperature vanishes like 
\begin{equation}
    T_c\approx D e^{-\frac{1}{\sqrt{\alpha^*-\alpha}}}.
\end{equation}

We can generalize our approach and consider a fermionic self energy in the normal
state that behaves as 
\begin{equation}
   \Sigma_{\boldsymbol{p}}\left(\epsilon\right)=-i\lambda{\rm sign}\left(\epsilon\right)\left|\epsilon\right|^{1-\gamma} 
\end{equation}
 with exponent  $0<\gamma<1$~\cite{Moon2010,Abanov2020}.  This allows putting our findings in more general context to describe different QCPs.
Examples are $\gamma=1/2$ for $d=2$ spin-density wave instabilities~\cite{Abanov2001,Abanov2003}
and $\gamma\rightarrow0^{+}$, i.e. $\Sigma\left(\epsilon\right)\sim\epsilon\log$$\epsilon$,
for $d=3$ color superconductivity due to gluon exchange~\cite{Son1999}
or for three-dimensional Ising-ferromagnetic spin fluctuations~\cite{Roussev2001,Chubukov2005}.
$\gamma\rightarrow1^{-},$ i.e. $\Sigma\left(\epsilon\right)\sim\log$$\epsilon$,
follows for $d=3$ massless bosons at very strong coupling\cite{Chubukov2005}.
$\gamma=1/3$ also describes composite fermions at half-filled Landau
levels~\cite{Bonesteel1996}, emergent gauge fields~\cite{Lee1989,Polchinski1994,Altshuler1994,Altshuler1995}
or nematic transitions~\cite{Halboth2000,Yamase2000} in two dimensions.
For each of these problems one can analyze a Hamiltonian similar
to Eq.\eqref{eq:Hamiltonian} within a related large-$N$ formulation.
While the case relevant to our problem corresponds to $\gamma=1/3$, we keep $0<\gamma <1$ arbitrary. Using this more general form of the self energy, the linearized gap-equation for the anomalous self energy  becomes 
\begin{equation}
\Phi\left(\epsilon\right)=\frac{\lambda_{p}}{\lambda}\int d\epsilon'\frac{\Phi\left(\epsilon\right)}{\left|\epsilon-\epsilon'\right|^{\gamma}\left|\epsilon'\right|^{1-\gamma}}.
\end{equation}
with the complement of the dimensionless coupling constant $\lambda$ of Eq.~\eqref{eq:lambda} in the pairing channel
\begin{equation}
    \lambda_{p}=\frac{1}{2}\lambda\left(1-\gamma\right)\left(1-\alpha\right).
    \label{eq:lambdap}
\end{equation}
The onset of pairing now occurs for $\alpha=\alpha^*$ where 
\begin{equation}
    1=\frac{\lambda_{p}(\alpha^*)}{\lambda}\int dx\frac{1}{\left|1-x\right|^{\gamma}\left|x\right|^{1-\gamma/2}}.
    \label{eq:pairing_cond_gamma}
\end{equation}
This recovers Eqs.~\eqref{eq:gamma-1} and \eqref{eq:pairing_cond_gamma_1/3} in the limit $\gamma=1/3$, as expected.

\section{Pairing Fluctuations}
\label{sec:Pairing Fluctuations}
 Next, we analyze the leading Gaussian
fluctuations of the action $S$ of Eq.~\eqref{eq:act} with respect to the
bilocal pairing fields $F$ and $\Phi$. This describes pairing fluctuations
of the critical normal state as well as the instability towards the onset of
superconductivity. It also lays the grounds for the derivation
of the holographic action which is the main goal of this paper. We perform our analysis at $T=0$, so strictly it  applies to the regime where the pair-breaking strength $\alpha$ is near the critical  $\alpha^{*}$. However, the description is expected to be applicable as long as the transition temperature is smaller than the temperature scale where Ising-ferromagnetic quantum critical fluctuations set in. Below we also discuss the extension of the holographic map to finite temperatures.

\begin{figure}
    \centering
    \includegraphics[width=0.9\linewidth]{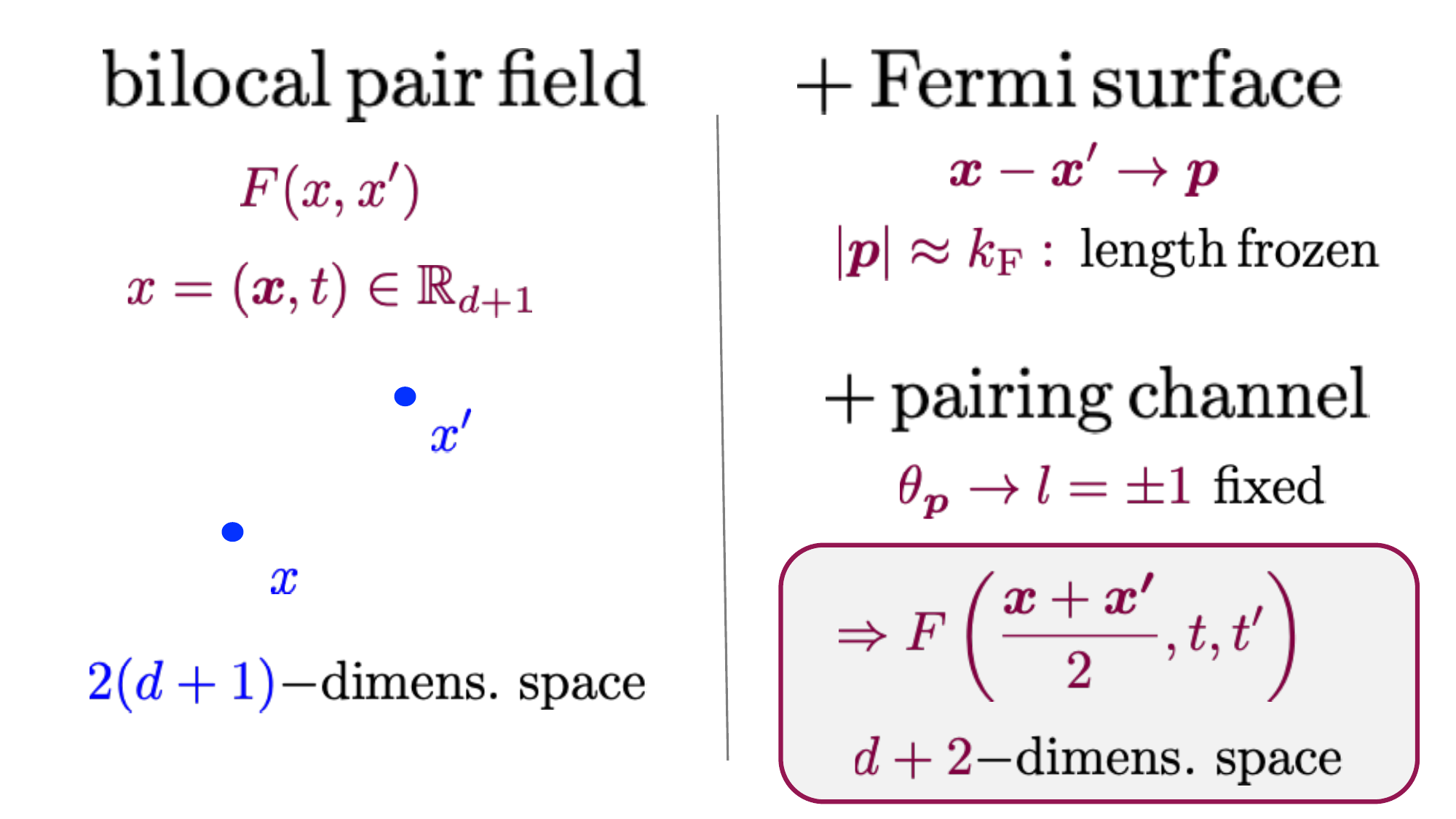}
    \caption{Illustration of the number of relevant coordinates of a bilocal pairing field in a compressible system. While the pair field $F(x,x')$, with $x$ and $x'$ $(d+1)$–component vectors, initially depends on $2(d+1)$ coordinates, the magnitude of the Fourier variable $\boldsymbol{p}$ associated with the relative spatial separation $\boldsymbol{x}-\boldsymbol{x}'$ is fixed by the Fermi surface. Its directional dependence is absorbed into the angular structure of the dominant pairing channel (a triplet in our case). This eliminates the dependence on the $d$ components of $\boldsymbol{p}$. As a result, the low–energy behavior depends effectively on only $2(d+1)-d=d+2$ coordinates.}
    \label{fig:dimensions}
\end{figure}

We expand the action Eq.~\eqref{eq:act} up to second order in pairing terms, i.e.
\begin{widetext}
\begin{eqnarray}
S_{sc}/N & = & \frac{1}{2} {\rm Tr}\left(\tilde{G}\otimes\hat{\Phi}^{\dagger}\otimes\hat{G}\otimes\hat{\Phi}\right)-\frac{1}{2}{\rm Tr}\left(\hat{F}^{\dagger}\otimes\hat{\Phi}+\hat{F}\otimes\hat{\Phi}^{\dagger}\right)
- \frac{\mu \check{g}_{p}^{2}}{2}\int_{x,x'}{\rm tr}\left(\hat{\sigma}^{z}\hat{F}\left(x,x'\right)\hat{\sigma}^{z}\hat{F}^{\dagger}\left(x',x\right)\right) D\left(x,x'\right).\nonumber
\end{eqnarray}
After Fourier transformation to momentum and frequency coordinates,
where $k$ refers to the variable conjugate to $(x+x')/2$ and $p$
conjugate to $x-x'$, we obtain:
\begin{eqnarray}
S_{sc}/N & = & -\frac{1}{2}\int_{kp}{\rm tr}\left(\hat{F}^{\dagger}\left(k,p\right)\hat{\Phi}\left(k,p\right)+h.c.\right)
 -  \frac{\mu \check{g}_{p}^{2}}{2}\int_{kpp'}{\rm tr}\left(\hat{\sigma}^z\hat{F}^{\dagger}\left(k,p\right)\hat{\sigma}^{z}\hat{F}\left(k,p'\right)\right) D\left(p-p'\right)\nonumber \\
 & - & \frac{1}{2}\int_{kp}{\rm tr}\left(\hat{G}\left(-\frac{k}{2}+p\right)\hat{\Phi}^{\dagger}\left(k,p\right)\hat{\Phi}\left(k,p\right)\hat{G}\left(-\frac{k}{2}-p\right) \right).
 \label{eq:SCaction_FT}
\end{eqnarray}
\end{widetext}
From the analysis of the gap equation we already know that the dependence
of the pairing function on the momentum $\boldsymbol{p}$ and on the
energy $\epsilon$ is very different. Hence, using
combined space-time variables $p=\left(\boldsymbol{p},\epsilon\right)$
or $k=\left(\boldsymbol{k},\omega\right)$ seizes to be efficient.
In what follows we expand the pairing functions
\begin{eqnarray}
\hat{F}\left(k,p\right) & = & \hat{F}_{\boldsymbol{k},\boldsymbol{p}}\left(\omega,\epsilon\right)=i\hat{\sigma}^{y}\hat{\sigma}^{j}\sum_{l}F_{\boldsymbol{k}l}\left(\omega,\epsilon\right)\eta_{\boldsymbol{p},l}
\end{eqnarray}
and same for $\hat{\Phi}\left(k,p\right)$, with respect to some
complete set of functions $\eta_{\boldsymbol{p},l}$
that only  depend on the direction of $\boldsymbol{p}$, i.e. on the angle $\theta_{\boldsymbol{p}}$.  The radial dependence is frozen by the Fermi surface, since $\left|\boldsymbol{p}\right| \approx k_{\rm F}$; see Fig.~\ref{fig:dimensions}. For an isotropic Fermi surface $l$ stands for the angular momentum quantum number. In the generic case we only assume that the set of functions is orthonormal
\begin{equation}
\left\langle l\mid l'\right\rangle =\int_{\boldsymbol{p}}\eta_{l}^{*}\left(\theta_{\boldsymbol{p}}\right)\eta_{l'}\left(\theta_{\boldsymbol{p}}\right)=\delta_{l,l'}
\end{equation}
where $\int_{\boldsymbol{p}}\cdots=k_{F}\int\frac{d\theta_{\boldsymbol{p}}}{2\pi v_{\boldsymbol{p}}}\cdots$
and $v_{\boldsymbol{p}}=\left|\boldsymbol{v}_{\boldsymbol{p}}\right|$
the magnitude of the Fermi velocity $\boldsymbol{v}_{\boldsymbol{p}}=\boldsymbol{v}\left(\theta_{\boldsymbol{p}}\right)$.
In the superconducting state, $\hat{F}(k,p)$ will condense in leading channel characterized by one of the $\eta_{\boldsymbol{p},l}$. However, for the moment we still need to perform the analysis of fluctuations in all channels and include the complete set $\{\eta_{\boldsymbol{p},l}\}$.

Inserting these expansions into Eq.~\eqref{eq:SCaction_FT}, we obtain for $j=1$ or $2$, i.e. the same triplet states discussed above, that
\begin{eqnarray}
S_{{\rm sc}}/N & = & \int\left(\Phi_{\boldsymbol{k}l}^{\dagger}\left(\omega,\epsilon\right)F_{\boldsymbol{k}l}\left(\omega,\epsilon\right)+\Phi_{\boldsymbol{k}l}\left(\omega,\epsilon\right)F_{\boldsymbol{k}l}^{\dagger}\left(\omega,\epsilon\right)\right)\nonumber \\
 & - & \mu \check{g}_{p}^{2}\int F_{\boldsymbol{k}l}^{\dagger}\left(\omega,\epsilon\right)D_{ll'}\left(\epsilon-\epsilon'\right)F_{\boldsymbol{k}l'}\left(\omega,\epsilon'\right)\nonumber \\
 & - & \int\Phi_{\boldsymbol{k}l}^{\dagger}\left(\omega,\epsilon\right)\chi_{\boldsymbol{k}ll'}\left(\omega,\epsilon\right)\Phi_{\boldsymbol{k}l'}\left(\omega,\epsilon\right),
\end{eqnarray}
where we introduced 
\begin{equation}
D_{ll'}\left(\epsilon\right)=2\int_{\boldsymbol{p},\boldsymbol{p}'}\eta_{\boldsymbol{p},l}^{*}\eta_{\boldsymbol{p}',l'}D_{\boldsymbol{p}-\boldsymbol{p}'}\left(\epsilon\right),
\end{equation}
 as well as
\begin{eqnarray}
\chi_{\boldsymbol{k},ll'}\left(\omega,\epsilon\right)&=&\int_{\boldsymbol{p}}\eta_{\boldsymbol{p},l}^{*}\hat{\eta}_{\boldsymbol{p},l'}  G_{\tfrac{\boldsymbol{k}}{2}-\boldsymbol{p}}\left(\tfrac{\omega}{2}-\epsilon\right)G_{\tfrac{\boldsymbol{k}}{2}+\boldsymbol{p}}\left(\tfrac{\omega}{2}+\epsilon\right).
\label{eq:chi_diagr}
\end{eqnarray}

Now we are in a position to integrate out the $\Phi_{\boldsymbol{k}l}\left(\omega,\epsilon\right)$
and obtain 
\begin{eqnarray}
S^{\left({\rm sc}\right)} & = & \int F_{\boldsymbol{k}l}^{\dagger}\left(\omega,\epsilon\right)\left(\chi_{\boldsymbol{k}}^{-1}\left(\omega,\epsilon\right)\right)_{ll'}F_{\boldsymbol{k}l'}\left(\omega,\epsilon\right)\nonumber \\
 & - & \mu g_{p}^{2}\int F_{\boldsymbol{k}l}^{\dagger}\left(\omega,\epsilon\right)D_{ll'}\left(\epsilon-\epsilon'\right)F_{\boldsymbol{k}l'}\left(\omega,\epsilon'\right).
 \label{eq:Gauss_from_SYK}
\end{eqnarray}
This formulation of the problem is particularly useful because the set of functions $\eta_{\boldsymbol{p},l}$ simultaneously diagonalizes both $\chi_{\boldsymbol{k}=0,ll'}\left(\omega=0,\epsilon\right)$ and $D_{ll'}\left(\epsilon\right)$ for all $\epsilon$.
However, at finite $\omega$ and $\boldsymbol{k}$ one must invert $\chi_{\boldsymbol{k},ll'}\left(\omega,\epsilon\right)$ in Eq.~\eqref{eq:chi_diagr}, which mixes the different modes and therefore requires retaining the complete set ${\eta_{\boldsymbol{p},l}}$. After performing this inversion in the limit of small $\omega$ and $\boldsymbol{k}$, we can then focus on the dominant pairing channel and work with the corresponding diagonal element $\chi^{-1}_{\boldsymbol{k},ll}\left(\omega,\epsilon\right)$.

For an isotropic Fermi surface follows 
\begin{equation}
D_{ll'}\left(\epsilon\right)=\delta_{ll'}\left(\frac{2}{3\sqrt{3}}\left(\frac{c^{2}}{2g^{2}\rho_{F}}\right)^{1/3}\left|\epsilon\right|^{-1/3}-\frac{\left|l\right|}{2}+\cdots\right).
\end{equation}
In the case of more generic Fermi surfaces, one still finds to leading order in frequency $D_{ll'}\left(\epsilon\right)\propto\delta_{ll'}\left|\epsilon\right|^{-1/3}$.
The corresponding analysis of the inverse particle-particle propagator is somewhat more tedious and given in Appendix~\ref{app:A}. The diagonal elements of the inverse $\chi^{-1}$, which are relevant for the leading instability, are given as
\begin{equation}
    \chi_{\boldsymbol{k}}^{-1}\left(\omega,\epsilon\right)=\frac{\lambda}{\pi}\left|\epsilon\right|^{1-\gamma}\left(1-\frac{\gamma\left(1-\gamma\right)}{8}\left(\frac{\omega}{\epsilon}\right)^{2}+\frac{\boldsymbol{k}^{2}}{k_{F}^{2}}\right).
    \label{eq:chi_pp}
\end{equation}
This is a rather surprising result. The  reason is that the last term, which governs the dependency on the total momentum $\boldsymbol{k}$ is formally sub-leading and not due to the universal low-energy modes of the problem.  By pure power counting, one would rather expect a behavior where $\boldsymbol{k}^{2}/k_{\rm F}^2$ is replaced by the more singular term $\left(\frac{E_{{\rm F}}}{\left|\epsilon\right|}\right)^{2\left(1-\gamma\right)}\boldsymbol{k}^{2}/k_{\rm F}^2$. While such terms occur at intermediate stages of the analysis, they exactly cancel in the final expression for $\chi_{\boldsymbol{k}}^{-1}\left(\omega,\epsilon\right)$. In Appendix~\ref{app:A} we show that this cancellation is valid for generic  Fermi surface geometries and is a generic feature for systems with a momentum-independent self energy. In the next section we will see that this cancellation has profound implications for the gravitational description of the problem. 


\section{Holographic Map} 
\label{sec:Holographic Map}

We are now in a position to establish the explicit map between the
Gaussian action Eq.~\eqref{eq:Gauss_from_SYK} in the leading pairing channel and the holographic theory of Eq.~\eqref{eq:HS_GL} in ${\rm AdS}_{2}\otimes\mathbb{R}_{2}$.
To this end we will proceed in two steps: First, we will summarize
some important properties of Euclidean ${\rm AdS}_{2}$, which is identical
to the two-dimensional hyperbolic space $\mathbb{H}_{2}$, and the
space $\mathbb{G}_{2}$ of the geodesics of $\mathbb{H}_{2}$. This space
of geodesics is also referred to as kinematic space~\cite{Czech2015,Boer2016}. In particular
we will discuss the non-local Radon transform $\tilde{\psi}={\cal R}\psi$
from a scalar field $\psi$, which is a complex function on $\mathbb{H}_{2}$,
while $\tilde{\psi}$ is a function on $\mathbb{G}_{2}$. In a second
step we establish a direct link between the fluctuating correlation
function $F_{\boldsymbol{k}}\left(\omega,\epsilon\right)$ for the
dominant pairing states $l=\pm1$ and a scalar field $\tilde{\psi}\left(\boldsymbol{k},\omega,z\right)$
that lives in kinematic space. Here $z\propto\left|\epsilon\right|^{-1}$
is an additional dimension that is sensitive to the internal dynamics
of the Cooper pair field. Combining these two steps we find that the
Yukawa-SYK theory that we start from exists in kinematic space and
allows for an explicit, albeit non-local relation to scalar fields
in the ${\rm AdS}_{2}\otimes\mathbb{R}_{2}.$

\subsection{Kinematic space and non-local Radon transform}

In this section we discuss the non-local relationship between functions
in Euclidean ${\rm AdS}_{2}\otimes\mathbb{R}_{2}$ and $\mathbb{G}_{2}\otimes\mathbb{R}_{2}$.
We use that Euclidean ${\rm AdS}_{2}$ is identical to the two-dimensional
hyperbolic space $\mathbb{H}_{2}$, while $\mathbb{G}_{2}$ is the
space of the geodesics of $\mathbb{H}_{2}$. Since the analysis of
this section is entirely local in the spatial coordinates $\boldsymbol{x}\in\mathbb{R}_{2}$
(or equivalently the two-dimensional momentum $\boldsymbol{k})$,
we suppress this dependency and merely analyze the kinematic space
and Radon transform of the $\mathbb{H}_{2}$ sector. Ref.~\cite{Stangier2026b} offers
a detailed for the more general case of the $D$-dimensional hyperbolic
space ${\rm \mathbb{H}_{D}}$; for a more general discussion of the
mathematical background of Radon transformations, see Ref.~\cite{Helgason2011}.
Detailed discussions  for $\mathbb{H}_{2}$ are given in Refs.~\cite{Das2018,Stone2025}

We start from the hyperboloid representation of the hyperbolic space
\begin{equation}
\mathbb{H}_{2}=\left\{ \vec{X}\in\mathbb{R}_{3}:c_{X}\left(\vec{X}\right)=-1\,\:{\rm and}\,\,X_{3}>0\right\} ,
\end{equation}
where $c_{X}\left(\vec{X}\right)=X_{1}^{2}+X_{2}^{2}-X_{3}^{2}.$
If we parametrize the $X_{i}\left(\xi^{\mu}\right)$ in terms of some
contra-variant coordinates, the induced metric follows as
\begin{equation}
g_{\mu\nu}=\sum_{ij=1}^{3}\eta_{ij}\frac{\partial X_{i}}{\partial\xi^{\mu}}\frac{\partial X_{j}}{\partial\xi^{\nu}}
\end{equation}
with $\eta_{ij}={\rm diag}\left(+1,+1,-1\right)$. A convenient set
of coordinates $\xi^{\mu}=\left(\tau,\zeta\right)$ is given by the
Poincare half plane 
\begin{equation}
\vec{X}=\left(\frac{\tau}{\zeta},\frac{\zeta}{2}\left(1+\frac{\tau^{2}-1}{\zeta^{2}}\right),\frac{\zeta}{2}\left(1+\frac{\tau^{2}+1}{\zeta^{2}}\right)\right),
\end{equation}
where $\zeta>0$. The induced metric then follows as 
\begin{equation}
ds^{2}=g_{\mu\nu}d\xi^{\mu}d\xi^{\nu}=\frac{d\tau^{2}+d\zeta^{2}}{\zeta^{2}},
\end{equation}
which is the Euclidean ${\rm AdS}_{2}$-part of Eq.~\eqref{eq:metric}. The analysis
of the geodesics of $\mathbb{H}_{2}$ is particularly convenient in
this hyperboloid representation as geodesic sub-manifolds are the
intersection of the hyperboloid with planes that pass through the
origin. These planes can be written as $\vec{X}\cdot\vec{G}=0$ and
are determined by the three-component vector $\vec{G}$. Changing
the length of $\vec{G}$ does not change the plane, i.e. $\left|\vec{G}\right|$
can be fixed by some arbitrary procedure and the geodesics are
determined by two parameters. Not all hyperplanes actually intersect
with the hyperboloid. It must hold that $c_{G}\left(\vec{G}\right)=-G_{1}^{2}-G_{2}^{2}+G_{3}^{2}<0$.
This allows to define the space of geodesics as
\begin{equation}
\mathbb{G}_{2}=\left\{ \vec{G}\in\mathbb{R}_{3}:c_{G}\left(\vec{G}\right)=-1\,\:{\rm and}\,\,G_{3}>0\right\} ,
\label{eq:DefG3}
\end{equation}
where $c_{G}\left(\vec{G}\right)=-1$ is used to fix the length of
$\vec{G}$. Each point in $\mathbb{G}_{2}$ then determines a one-dimensional
geodesic sub-manifolds of $\mathbb{H}_{2}$. Similar to $\mathbb{H}_{2}$,
we can use a set of coordinates $\tilde{\xi}^{\mu}=\left(t,z\right)$
to parametrize $\mathbb{G}_{2}$, where
\begin{equation}
\vec{G}=\left(\frac{t,}{z},\frac{z}{2}\left(-1+\frac{t^{2}-1}{z^{2}}\right),\frac{z}{2}\left(1-\frac{t^{2}+1}{z^{2}}\right)\right)
\label{eq:ParamG3}
\end{equation}
 ensures that $\vec{G}\in\mathbb{G}_{2}$ as defined above, with $t\in \mathbb{R}$ and $z>0$~\footnote{The condition $G_{3}>0$ in Eq.~\eqref{eq:DefG3} is not satisfied by the parametrization of Eq.~\eqref{eq:ParamG3} for all  $t\in \mathbb{R}$ and $z>0$.
However, this condition is somewhat arbitrary and serves only to ensure that the geodesics of a given plane are not counted twice, in view of the overall sign ambiguity of $\vec{G}$. Hence, any alternative partition of the space into two halves - such that each point in one half can be obtained from a point in the other by reversing the sign of $\vec{G}$ - would work equally well.
The parametrization of Eq.~\eqref{eq:ParamG3} rather corresponds to the partition  $G_{3}<-G_{2}$ instead of  $G_{3}>0$.}. The condition
$\vec{X}\cdot\vec{G}=0$ can now easily be written as 
\begin{equation}
z^{2}=\zeta^{2}+\left(\tau-t\right)^{2}.
\end{equation}
The geodesics are semicircles of radius $z$ and centered around $\left(0,t\right)$.
The induced metric of $\mathbb{G}_{2}$, which is determined in full analogy
to the one for $\mathbb{H}_{2}$, is given as
\begin{equation}
d\tilde{s}^{2}=h_{\mu\nu}d\tilde{\xi}^{\mu}d\tilde{\xi}^{\nu}=\frac{dz^{2}-dt^{2}}{z^{2}}.
\end{equation}
This is the metric of de Sitter space, where we merely use a convention with a different overall minus sign.

The close relationship between $\mathbb{H}_{2}$ and its space of
geodesics suggests to perform a non-local integral transform from
functions defined on one space to functions on the other. This is
accomplished by the Radon transformation 
\begin{equation}
\tilde{\psi}={\cal R}\psi,
\end{equation}
where ${\cal R}$ corresponds to the integration over the ${\rm AdS}_{2}$
geodesics, parametrized by the coordinates $\tilde{\xi}$: 
\begin{equation}
\tilde{\psi}\left(\tilde{\xi}\right)=\int_{\tilde{\xi}}d\lambda\phi\left(\xi\left(\lambda\right)\right).
\end{equation}
An important aspect of Radon transforms is the intertwinement of the
Laplacians before and after the transformation~\cite{Helgason2011,Stangier2026b}:
\begin{equation}
\Box_{{\rm \mathbb{G}_{2}}}{\cal R}\phi={\cal R}\left(\Box_{{\rm \mathbb{H}_{2}}}\phi\right).\label{eq:Laplacians}
\end{equation}
This intertwinement of the Laplacians holds, in particular, for the
normalized eigenfunctions $\eta_{p}$ of $\Box_{{\rm \mathbb{H}_{2}}}$
with eigenvalue $-p^{2}-\frac{1}{4}$ which we express relative to
the Breitenlohner Freedman bound. Since ${\cal R}\left(\Box_{{\rm \mathbb{H}_{2}}}\eta_{p}\right)=-\left(p^{2}+\frac{1}{4}\right){\cal R}\eta_{p}$,
it follows that $\Box_{{\rm \mathbb{G}_{2}}}{\cal R}\eta_{p}=-\left(p^{2}+\frac{1}{4}\right){\cal R}\eta_{p}.$
The Radon transform of an eigenfunction is itself eigenfunction with
same eigenvalue. It is however not guaranteed, that the ${\cal R}\eta_{p}$
is properly normalized. In Refs.~\cite{Das2018,Stone2025} it was shown that 
\begin{equation}
{\cal R}\eta_{p}=L_{p}\tilde{\eta}_{p}\label{eq:Leg factors},
\end{equation}
where $\tilde{\eta}_{p}$ are the properly normalized eigenfunctions
of $\Box_{{\rm \mathbb{G}_{2}}}$. The so-called leg factors are given
as
\begin{equation}
L_{p}=-2i\sqrt{\pi}\frac{\Gamma\left(\frac{1}{4}+i\frac{p}{2}\right)}{\Gamma\left(\frac{3}{4}+i\frac{p}{2}\right)}.
\end{equation}
The intertwinement of the Laplacians can also be used to show that the Radon transformation can always be inverted~\cite{Stangier2026b}.

These insights allow us to establish a relation between the action
of a scalar field in Euclidean ${\rm AdS}_{2}$ 
\begin{equation}
S=\int_{\mathbb{H}_{2}}d^{2}\xi\sqrt{g}\psi^{*}\left(m^{2}-\Box_{{\rm \mathbb{H}_{2}}}\right)\psi\label{eq:act_H2}
\end{equation}
and its counterpart in $\mathbb{G}_{2}$
\begin{equation}
\tilde{S}=\int_{\mathbb{G}_{2}}d^{2}\tilde{\xi}\sqrt{-h}\tilde{\psi}^{*}\left(\tilde{m}^{2}-\Box_{{\rm \mathbb{G}_{2}}}\right)\tilde{\psi}.\label{eq:act_G2}
\end{equation}
To this end, we first expand both fields in terms of the normal modes,
i.e. $\psi\left(\xi\right)=\int dpa_{p}\eta_{p}\left(\xi\right)$
and $\tilde{\psi}\left(\tilde{\xi}\right)=\int dp\tilde{a}_{p}\tilde{\eta}_{p}\left(\tilde{\xi}\right).$
Inserting these in the respective actions yields
\begin{eqnarray}
S & = & \int dp\left(m^{2}+p^{2}+\frac{1}{4}\right)\left|a_{p}\right|^{2}\label{eq:expansion AdS2}
\end{eqnarray}
and 
\begin{eqnarray}
\tilde{S} & = & \int dp\left(\tilde{m}^{2}+p^{2}+\frac{1}{4}\right)\left|\tilde{a}_{p}\right|^{2}.\label{expansion dS2}
\end{eqnarray}
We can now Radon-transform the expansion for $\psi$ in normal modes.
Together with the leg factors of Eq.(\ref{eq:Leg factors}) follows
that the expansion coefficients are related by
\begin{equation}
\tilde{a}_{p}=L_{p}a_{p}.
\end{equation}
Thus, we obtain
\begin{equation}
\tilde{S}=\int dp\left(\tilde{m}^{2}+p^{2}+\frac{1}{4}\right)\left|L_{p}\right|^{2}\left|a_{p}\right|^{2}.
\end{equation}
At small $p$ holds $\left|L_{p}\right|^{2}=\frac{4\pi\Gamma\left(\frac{1}{4}\right)^{2}}{\Gamma\left(\frac{3}{4}\right)^{2}}\left(1-4Gp^{2}\right)$,
where $G=\sum_{n=0}^{\infty}\frac{\left(-1\right)^{n}}{\left(2n+1\right)^{2}}\approx0.915966$
is Catalan's constant. Hence, it follows 
\begin{equation}
S\left[m,\psi\right]=\kappa\tilde{S}\left[\tilde{m},{\cal R}\psi\right],\label{eq:action_to_action}
\end{equation}
with overall coefficient $\kappa^{-1}=4\pi\Gamma\left(\frac{1}{4}\right)^{2}\left[1-4G\left(m^{2}+\frac{1}{4}\right)\right]$
and
\begin{equation}
m^{2}=\frac{\tilde{m}^{2}+\frac{1}{4}}{\left[1-4G\left(\tilde{m}^{2}+\frac{1}{4}\right)\right]}-\frac{1}{4}.
\label{eq:m_vs_mtilde}
\end{equation}
At the BF bound $m^{2}=\tilde{m}^{2}$ while deviations between $m^{2}$ and $\tilde{m}^{2}$ are small throughout, such that the two actions
are, up to a numerical constant, the same.
Eq.~\eqref{eq:action_to_action},
is the key result of this section. It demonstrates that a problem
in kinematic space, i.e. the space of geodesics, that is governed
by the action Eq.~\eqref{eq:act_G2}, can be non-locally mapped to a problem
in Euclidean ${\rm AdS}_{2}$  and that is governed by Eq.~\eqref{eq:act_H2}. 
Next we will show that the action Eq.~\eqref{eq:Gauss_from_SYK} is in fact equivalent to the theory in kinematic space.

\subsection{Local map to the kinematic space}

We continue our analysis of the Gaussian action of Eq.~\eqref{eq:Gauss_from_SYK} which
we evaluate in the dominant pairing channel ($l=\pm1$ for a rotationally invariant Fermi surface). For simplicity,
we drop the index $l$ from now on. Couplings between the $l=+1$
and $l=-1$ channel, or the components of a higher-dimensional irreducible representation of a point group,  do not occur at the Gaussian level. Going beyond
the Gaussian regime, such couplings give rise to the usual behavior
of a two-component order parameter~\cite{Sigrist1991}.

Thus, we start from
\begin{eqnarray}
S^{\left({\rm sc}\right)} & = & \int\frac{d^{2}kd\omega d\epsilon}{\left(2\pi\right)^{4}}F_{\boldsymbol{k}}^{\dagger}\left(\omega,\epsilon\right)\chi_{\boldsymbol{k}}^{-1}\left(\omega,\epsilon\right)F_{\boldsymbol{k}}\left(\omega,\epsilon\right)\nonumber \\
 & - & 2\lambda_{p}\int\frac{d^{2}kd\omega d\epsilon d\epsilon'}{\left(2\pi\right)^{5}}\frac{F_{\boldsymbol{k}}^{\dagger}\left(\omega,\epsilon\right)F_{\boldsymbol{k}}\left(\omega,\epsilon'\right)}{\left|\epsilon-\epsilon'\right|^{\gamma}},\label{eq:act_orig_lv}
\end{eqnarray}
with $\chi_{\boldsymbol{k}}^{-1}\left(\omega,\epsilon\right)$ of
Eq.~\eqref{eq:chi_pp} and the $\lambda_p$ the coupling constant in the pairing channel as given in Eq.~\eqref{eq:lambdap}. The fluctuating anomalous correlation function $F_{\boldsymbol{k}}\left(\omega,\epsilon\right)$
depends on the two component of the spatial momentum $\boldsymbol{k}=\left(k_{x},k_{y}\right)$
as well as the frequencies that are conjugate to the absolute and
relative times, $\omega$ and $\epsilon$, respectively. Hence it
lives four dimensions, i.e. one extra dimension compared to the $2+1$
dimensional space time we usually use for a two-dimensional quantum
system; see Fig.~\ref{fig:dimensions}. We now  rewrite $F_{\boldsymbol{k}}\left(\omega,\epsilon\right)$ as
\begin{equation}
F_{\boldsymbol{k}}\left(\omega,\epsilon\right)=c_{0}\epsilon^{\frac{\gamma-1}{2}}\tilde{\psi}\left(k,c/\epsilon\right),\label{eq:local_map}
\end{equation}
where the constants $c_{0}=\sqrt{\frac{\pi}{2\lambda_{p}bc}}$ and
$c=\sqrt{\gamma\left(1-\gamma\right)/8}$ are chosen to obtain convenient
expressions in terms of the field $\tilde{\psi}$. We further used
$k=\left(\boldsymbol{k},\omega\right)$ for the $2+1$-dimensional
space-time momenta and will denote $z=c/\left|\epsilon\right|$. 

\begin{figure}
    \centering
    \includegraphics[width=0.9\linewidth]{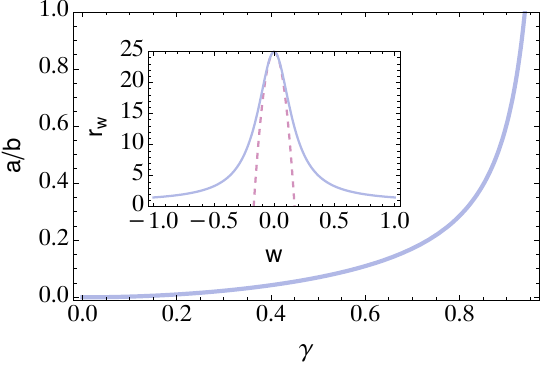}
    \caption{Ratio $a/b$  that determines the low-energy behavior $r_w=a-bw^2\cdots $ as function of the exponent $\gamma$.  $r_w$ is given in  Eq.~\eqref{eq:r_w}. It diagonalizes the $S_2$ part of the action after Mellin transformation. The inset shows $r_w$ (solid line) along with the  quadratic expansion at small $w$ for $\gamma=1/3$ (dashed line). }
    \label{fig:aoverb}
\end{figure}
We analyze the two terms in Eq.~\eqref{eq:act_orig_lv} under
the transformation Eq.~\eqref{eq:local_map} separately. The first term
follows immediately and is given as
\begin{eqnarray}
S_{1} & = & \int\frac{d^{2}kd\omega d\epsilon}{\left(2\pi\right)^{4}}\chi_{\boldsymbol{k}}^{-1}\left(\omega,\epsilon\right)\left| F_{\boldsymbol{k}}\left(\omega,\epsilon\right)\right|^2\nonumber \\
 & = & \frac{\lambda}{\lambda_{p}b}\int\frac{d^{2}kd\omega dz}{\left(2\pi\right)^{4}}\left(\frac{1}{z^{2}}-\omega^{2}+\frac{\boldsymbol{k}^{2}}{z^2 k_{F}^{2}}\right)\left| \tilde{\psi}\left(k,z\right)\right|^2.
 \label{eq:rewrittenS1}
\end{eqnarray}
The second part
\begin{equation}
S_{2}=-2\lambda_{p}\int\frac{d^{2}kd\omega d\epsilon d\epsilon'}{\left(2\pi\right)^{5}}\frac{F_{\boldsymbol{k}}^{\dagger}\left(\omega,\epsilon\right)F_{\boldsymbol{k}}\left(\omega,\epsilon'\right)}{\left|\epsilon-\epsilon'\right|^{\gamma}}
\end{equation}
is slightly more subtle, as we have to perform an additional gradient
expansion, valid at low energies. 
Since $S_{2}$ is local in $\boldsymbol{k}$ and $\omega$, it is
sufficient and  less cumbersome to write $S_{2}=\int\frac{d^{2}kd\omega}{\left(2\pi\right)^{3}}s_{2}\left(k\right)$
and consider 
\begin{equation}
s_{2}=-2\lambda_{p}\int\frac{d\epsilon d\epsilon}{\left(2\pi\right)^{2}}\frac{F\left(\epsilon\right)F\left(\epsilon'\right)}{\left|\epsilon-\epsilon'\right|^{\gamma}},
\label{eq:pairin_int_s2}
\end{equation}
where we suppress the dependency on $k$ for the moment. We first
notice that $s_{2}$ can be diagonalized by a  Fourier transform of logarithmic variables, i.e. a Mellin transform 
\begin{equation}
F\left(\epsilon\right)=\int_{-\infty}^{\infty}dwf_{w}\left|\epsilon\right|^{-iw-1+\gamma/2}.
\end{equation}
In terms of $f_{w}$
follows
\begin{equation}
s_{2}=-\frac{2\lambda_{p}}{\pi}\int dwr_{w}\left|f_{w}\right|^{2},
\end{equation}
with 
\begin{eqnarray}
r_{w} & = & \int_{-\infty}^{\infty}dx\frac{\left|x\right|^{iw}}{\left|1-x\right|^{\gamma}\left|x\right|^{1-\gamma/2}}\nonumber \\
 & = & \frac{\frac{1}{2}\pi^{2}}{\cos\left(\frac{\pi\gamma}{2}\right)\Gamma\left(\gamma\right)\left|\Gamma\left(1+iw-\frac{\gamma}{2}\right)\sinh\left(\frac{\left(2w-i\gamma\right)\pi}{4}\right)\right|^{2}}.
 \label{eq:r_w}
\end{eqnarray}
At small $w$ follows $r_{w}\approx a-bw^{2}$ where $a$ and $b$
are both positive for $0<\gamma<1$. In Fig.~\ref{fig:aoverb} we show $r_w$ for $\gamma=1/3$ in the inset, along with the quadratic expansion for small $w$. The figure also shows the ratio $a/b$ of the two expansion coefficients. 
Notice, the expansion of $r_{w}$ with respect to $w$ is not
a gradient expansion in $\partial_{z}\tilde{\psi}$. Instead, it is an
expansion for small $\left(z\partial_{z}-\frac{1}{2}\right)\tilde{\psi}$
which is the appropriate expansion for the curved space under consideration.
This expansion allows us to write at small
$w$ 
\begin{equation}
s_{2}\approx\frac{2\lambda_{p}}{\pi}\int dw\left(-a+bw^{2}\right)\left|f_{w}\right|^{2}.
\end{equation}
If we furthermore define the Mellin transform 
\begin{equation}
\tilde{\psi}\left(z\right)=\int_{-\infty}^{\infty}ds\tilde{\psi}_{s}z^{-is+\frac{1}{2}},
\end{equation}
we can express the local map Eq.~\eqref{eq:local_map} as $f_{w}=c_{0}c^{iw+\frac{1}{2}}\tilde{\psi}_{-w}$ and find
\begin{equation}
s_{2}=\int\frac{ds}{\pi}\left(-\frac{a}{b}+s^{2}\right)\left|\tilde{\psi}_{s}\right|^{2}
\end{equation}
in terms of $\tilde{\psi}$. The inverse Mellin transform is easily
found to be $\tilde{\psi}_{s}=\int_{0}^{\infty}\frac{d\zeta}{2\pi}\tilde{\psi}\left(\zeta\right)\zeta^{is-\frac{3}{2}}$
and  the non-local action of Eq.~\eqref{eq:pairin_int_s2} can be expressed in terms of $\tilde{\psi}$: 
\begin{equation}
s_{2}=\int_{0}^{\infty}\frac{dz}{2\pi}\left(\left|\partial_{z}\tilde{\psi}\right|^{2}-\left(\frac{1}{4}+\frac{a}{b}\right)\frac{1}{z^{2}}\left|\tilde{\psi}\right|^{2}\right).
\label{eq:rewrittenS2}
\end{equation}

\subsection{The explicit holographic map}
We are now in a position to rewrite the full Gaussian action of Eq.(\ref{eq:act_orig_lv}), by combining the result of Eq~\eqref{eq:rewrittenS1} for $S_1$ with Eq.~\eqref{eq:rewrittenS2}  for $S_2$. This leads to the collective action of superconducting fluctuations in terms of $\tilde{\psi}$:
\begin{eqnarray}
S_{\rm sc} & = & \int\frac{d^{2}kd\omega dz}{\left(2\pi\right)^{4}}\left[\left|\partial_{z}\tilde{\psi}\right|^{2}+\left(\frac{\tilde{m}^{2}}{z^{2}}-\omega^{2}+\frac{\boldsymbol{k}^{2}}{z^2 k_{F}^{2}}\right)\left|\tilde{\psi}\right|^{2}\right]
\label{eq:act_kinem}
\end{eqnarray}
with mass
\begin{equation}
\tilde{m}^{2}=-\frac{1}{4}+\frac{1}{b}\left(\frac{\lambda}{\lambda_{p}}-a\right).
\label{eq:mtil}
\end{equation}
Fourier transformation from momentum and frequency to position
and time  space, $\left(\boldsymbol{k},\omega\right)\rightarrow\left(\boldsymbol{x},t\right)$, allows to write this result as in the geometric form
\begin{equation}
S_{\rm sc} =\int d^{4}\tilde{\xi}\sqrt{-h}\left(\partial^{\mu}\tilde{\psi}^{*}\partial_{\mu}\tilde{\psi}+\tilde{m}^{2}\left|\tilde{\psi}\right|^{2}\right)\label{eq:dS_R2}
\end{equation}
with four-dimensional coordinates $\tilde{\xi}^{\mu}=\left(\boldsymbol{x},t,z\right)$ and metric
\begin{equation}
d\tilde{s}^{2}=h_{\mu\nu}d\tilde{\xi}^{\mu}d\tilde{\xi}^{\nu}=\frac{dz^{2}-dt^{2}}{z^{2}}+k_{F}^{2}d\boldsymbol{x}^{2},\label{eq:metricdS_R2}
\end{equation}
where $h={\rm det}h_{\mu\nu}$. Given the analysis of the previous section, we recognize that this metric describes the kinetic space $\mathbb{G}_{2}\otimes\mathbb{R}_{2}$ of ${\rm AdS}_{2}\otimes\mathbb{R}_{2}$. Performing the Radon transform at each point $\boldsymbol{x}$ implies that we obtain the holographic action of a Cooper-pair field in ${\rm AdS}_{2}\otimes\mathbb{R}_{2}$.
\begin{figure}
    \centering
    \includegraphics[width=0.95\linewidth]{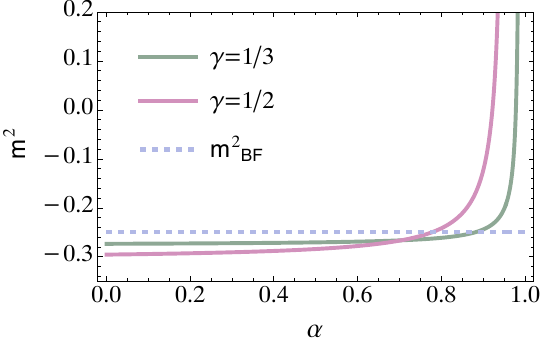}
    \caption{Mass $m^2$ of the scalar Cooper-pair field in ${\rm AdS}_2 \otimes \mathbb{R}_2$ for two values of the exponent governing the quantum-critical self-energy, shown as a function of the pair-breaking parameter $\alpha$. The dashed line indicates the threshold mass $m^2_{\rm BF}$, with pairing occurring for $m^2 < m^2_{\rm BF}$. The chosen values of $\gamma$ are relevant to the Ising ferromagnetic transition ($\gamma = 1/3$) and to spin-density-wave criticality ($\gamma = 1/2$). }
    \label{fig:masses}
\end{figure}

The resulting explicit nonlocal map can be conveniently expressed in momentum–frequency space and establishes a direct connection between the Gor'kov function $F_{\boldsymbol{k}}\left(\omega,\epsilon\right)$, arising in a fully dynamical and spatially inhomogeneous  theory of superconductivity, and the scalar field $\psi$ of the holographic superconductor in $\mathbb{H}_2\otimes \mathbb{R}_2$, as it appears in Eq.~\eqref{eq:HS_GL}:
\begin{eqnarray}
F_{\boldsymbol{k}}\left(\omega,\epsilon\right)&=&2c_{0}c\int_{0}^{c\left|\epsilon\right|^{-1}}d\zeta\frac{\cos\left(\omega\sqrt{\left(c/\epsilon\right)^{2}-\zeta^{2}}\right)}{\left|\epsilon\right|^{\frac{3-\gamma}{2}}\zeta\sqrt{\left(c/\epsilon\right)^{2}-\zeta^{2}}}
\nonumber \\
&\times & \psi\left(\boldsymbol{k},\omega,\zeta\right).
\label{eq:the_map}
\end{eqnarray}
The mass $m$ in Eq.~\eqref{eq:HS_GL} is related via Eq.~\eqref{eq:m_vs_mtilde} to $\tilde{m}$ of Eq.~\eqref{eq:mtil}.   In Fig.~\ref{fig:masses} we show the dependency of $m^2$ as function of the pair-breaking parameter $\alpha$ for the two cases of $\gamma=1/3$ and $\gamma=1/2$. It illustrates the fact that we can determine the properties  of the holographic theory from microscopic parameters of our initial Hamiltonian.

Eq.~\eqref{eq:the_map} constitutes the key result of this paper. This reformulation makes explicit the geometric structure underlying the initial power-law behavior of the pairing susceptibility $\chi_{\boldsymbol{k}}\left(\omega,\epsilon\right)$ and the pairing interaction $\propto\left|\epsilon-\epsilon'\right|^{-\gamma}$. It also demonstrates that the spatial sector remains flat and is therefore unaffected by curvature.
By contrast, the curved part of the space corresponds, in the original field-theory description, to the space of geodesics of ${\rm AdS_{2}}$. This correspondence necessitates a non-local map between the field theory and its holographic formulation.

\begin{figure}
    \centering
    \includegraphics[width=0.95\linewidth]{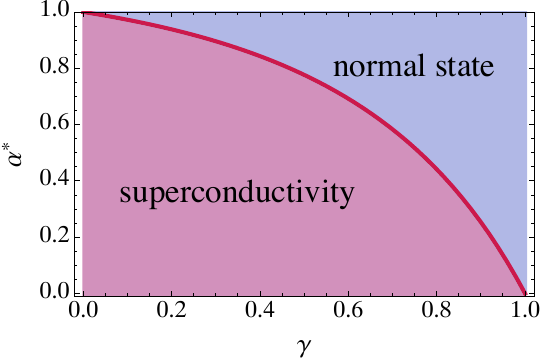}
    \caption{Critical pair-breaking strength $\alpha^*$ at which superconductivity disappears - as determined from Eq.~\eqref{eq:pairing_cond_hol}, i.e. $m^2=m^2_{\rm BF}$ - shown as a function of the exponent $\gamma$ that controls the frequency dependence of the self-energy. The superconducting state is robust for small $\gamma$, but becomes increasingly fragile as $\gamma \rightarrow 1$.   }
    \label{fig:PD}
\end{figure}

A first application of this map is to identify the onset of superconductivity at zero temperature. Within the holographic formalism this corresponds to case where the $m=m_{\rm BF}$, where $m^2_{\rm BF}=-\frac{1}{4}$ corresponds to the Breitenlohner and
Freedman bound of the mass~\cite{Breitenlohner1982}. Using  $m\rightarrow \tilde{m}$ near this bound and Eq.~\eqref{eq:mtil} yields the condition
\begin{equation}
    1=\frac{\lambda_{p}}{\lambda}a=\frac{\lambda_{p}}{\lambda}\int_{-\infty}^{\infty}dx\frac{1}{\left|1-x\right|^{\gamma}\left|x\right|^{1-\gamma/2}},
    \label{eq:pairing_cond_hol}
\end{equation}
where we used that $a=r_{w=0}$ with $r_w$ of Eq.~\eqref{eq:r_w}.
This is precisely the same condition as what was obtained in Eq.~\eqref{eq:pairing_cond_gamma} from the solution of the Eliashberg theory and demonstrates that both perspectives, quantum critical Eliashberg theory and holographic superconductivity are indeed identical for the compressible two-dimensional systems discussed here. The resulting phase diagram, illustrating the transition from the superconducting to the normal state, is shown in Fig.~\ref{fig:PD}.

Because of the ${\rm AdS}_{2}\otimes\mathbb{R}_{2}$ structure of the holographic theory, it was possible to use many of the technical steps of the holographic map that was used in Ref.~\cite{Inkof2022} for the zero-dimensional Yukawa-SYK model with an ${\rm AdS}_{2}$ gravity formulation. It furthermore allows us to use the same conformal maps to relate the zero-temperature theory to the one at finite temperature, both in the field theory formulation and in the gravitational theory. An important implication is that the metric that one obtains at finite $T$ is then given as
\begin{equation}
    ds^{2}=\frac{f\left(\zeta\right)d\tau^{2}+f\left(\zeta\right)^{-1}d\zeta^{2}}{\zeta^{2}}+k_{F}^{2}d\boldsymbol{x}^{2},
\end{equation}
with blackening factor $f\left(\zeta\right)=1-\zeta^{2}$/$\zeta_{T}^{2}$
that changes its sign at the horizon $\zeta_{T}^{-1}=2\pi T$. Hence,
the ${\rm AdS}_{2}$ sector is  governed by a black hole that
 signals that the scaling of the quantum critical theory
stops when energies are comparable to temperature.  This black-hole formulation directly follows from the Hamiltonian of Eq.~\eqref{eq:Hamiltonian}. For the explicit expression of this map, that replaces Eq.~\eqref{eq:the_map} at finite $T$, see Ref.~\cite{Inkof2022}.

In Eq.\eqref{eq:HS_GL} we have no gauge field in the gravitational
bulk, i.e $\partial_{\zeta}$ instead of $\partial_{\zeta}-i2eA_{\zeta}$
with corresponding electric field ${\cal E}$. This is consistent
with the holographic map of the Yukawa-SYK model where one finds ${\cal E}=0$
at particle-hole symmetry\cite{Inkof2022}, a symmetry that is emergent
for a system with Fermi surface.

\subsection{Reissner-Nordstr\"om  versus Lifshitz gravity}
The ${\rm AdS}_{2}\otimes\mathbb{R}_{2}$ geometry that emerges in our analysis indicates that the dynamical sector is effectively local, and that spatial fluctuations do not play a significant role in determining the properties of the Cooper pair field. Within  many-body theory, this reflects the absence of any singular momentum dependence in the fermionic self-energy. Although the ferromagnetic fluctuations responsible for the onset of superconductivity are characterized by a diverging length scale, the fermionic excitations themselves exhibit a behavior that is singular in time but local in space. 
Within a gravitational perspective this corresponds to the behavior in the vicinity of the horizon of a Reissner-Nordstr\"om black hole, a geometry that also factorizes at low energies into ${\rm AdS_{2}\otimes\mathbb{R}_{d}}$~\cite{Sachdev2012,Zaanen2015,Ammon2015,Hartnoll2018,Baggioli2019}. Here correlations also become completely local in space and non-local only in time\cite{Faulkner2011,Iqbal2012}.

As we discussed, the ${\rm AdS}_{2}\otimes\mathbb{R}_{2}$ geometry is a consequence of the cancellation of terms in the pairing response of Eq.~\eqref{eq:chi_pp}. Let us for the moment consider the case where this cancellation did not take place. Then, we would have, instead of the momentum dependence $\boldsymbol{k}^{2}/k_{\rm F}^2$  in Eq.~\eqref{eq:chi_pp}, a term that behaves like $\left(\frac{E_{{\rm F}}}{\left|\epsilon\right|}\right)^{2\left(1-\gamma\right)}\boldsymbol{k}^{2}/k_{\rm F}^2$. This behavior directly follows from power counting. Such a term would then dominate the low-energy behavior and change our entire analysis. If one then repeats the analysis that determines the action in kinematic space, i.e. that leads to Eq.~\eqref{eq:act_kinem}, one finds instead the line element
\begin{equation}
d\tilde{s}^{2}=\frac{dz^{2}-dt^{2}}{\zeta^{2}}-\frac{d\boldsymbol{x}^{2}}{\zeta^{2/z_{\rm ds}}}.\label{eq:metric_Lif}
\end{equation}
This metric describes a Lifshitz geometry\cite{Kachru2008,Taylor2008,Gubser2009,Hartnoll2010,Hartnoll2011,Hartnoll2011b,Huijse2012,Liu2013,Taylor2016}, albeit in de Sitter space that is expected to describe the corresponding kinematic space. The  dynamic scaling exponent of this geometry is $z_{\rm ds}=\left(1-\gamma\right)^{-1}$. To see this interpretation of $z_{\rm ds}$ explicitly, rescale $z\rightarrow s z$,
$t\rightarrow s t$ and $x_{i}\rightarrow s^{1/z_{\rm ds}}x_{i}$ with 
parameter $s$, which ensures invariance of the line element.  For the case of the Ising ferromagnetic critical point, this implies $z_{{\rm ds}}=3/2$. Hence, the natural dynamic scaling exponent of the holographic theory
is not $z^{\rm bos}_{{\rm ds}}=3$~\cite{
Hertz1976,Millis1993} that follows from balancing the boson momentum and energy. Indeed, balancing $\Sigma\left(\epsilon\right)\sim\left|\epsilon\right|^{1-\gamma}$
against $v_{F}p_{\perp}$ leads to $\epsilon\sim p_{\perp}^{z_{\rm ds}}$ with
$z_{\rm ds}$ given above.

It is certainly of interest to identify microscopic theories that generate such a Lifshitz gravity. In this context we mention that in Ref.~\cite{Stangier2026b} we show that Dirac problems do realize an emergent gravity theory where time and space are both part of a non-trivial geometry. In this case we find a metric like Eq.~\eqref{eq:metric_Lif}, however with dynamic scaling exponent $z_{\rm ds}=1$ which yields ${\rm AdS}_{4}$.

\section{Conclusion}
In this work we derived a microscopic holographic formulation of superconductivity in a compressible quantum-critical metal with a critical Fermi surface. Starting from a two-dimensional large-$N$, $M$ Yukawa–Sachdev-Ye-Kitaev model of fermions coupled to Ising-ferromagnetic fluctuations, we reformulated the pairing problem in terms of bilocal collective fields and analyzed Gaussian pairing fluctuations of the quantum-critical normal state. This allowed us to establish an explicit mapping between the Eliashberg theory of quantum-critical pairing and a scalar field theory in an emergent curved spacetime with ${\rm AdS}_{2}\otimes\mathbb{R}_{2}$ geometry. 

A central result of our analysis is that the additional holographic dimension encodes the internal temporal dynamics of fluctuating Cooper pairs. The map between the anomalous Gor’kov function and the scalar field in the gravitational description is intrinsically nonlocal and can be formulated in terms of a Radon transform relating fields in AdS$_2$ to fields in its kinematic space, i.e. its space of geodesics. Within this framework, the superconducting instability is naturally interpreted as a Breitenlohner–Freedman instability of the scalar field. We demonstrated that this geometric criterion is exactly equivalent to the onset of pairing obtained from the linearized Eliashberg equations, thereby providing a direct microscopic connection between holographic superconductivity and quantum-critical pairing in metals. 

The factorized ${\rm AdS}_{2}\otimes\mathbb{R}_{2}$ geometry reflects a key physical property of the underlying quantum-critical state: while magnetic fluctuations exhibit a diverging spatial correlation length, fermionic excitations remain local in space but display singular temporal dynamics. This locality in momentum space leads to a cancellation of more singular gradient terms in the pairing susceptibility and is ultimately responsible for the emergence of an AdS$_2$ sector rather than a Lifshitz geometry. From the gravitational perspective, this behavior is closely related to the near-horizon structure of a Reissner–Nordström black hole, which captures the physics of compressible quantum-critical matter. 

Our results extend earlier derivations of holographic superconductivity from zero-dimensional SYK-type models to systems with a  Fermi surface and spatial structure. They therefore provide a concrete microscopic foundation for the use of holographic methods in strongly correlated metallic systems. The present formulation also suggests several directions for future work. An important extension is the analysis of pairing beyond the Gaussian regime, where nonlinear couplings between different pairing channels determine the symmetry of the ordered state. It will also be interesting to investigate transport and collective modes within the holographic framework derived here, as well as the interplay between superconductivity and competing instabilities near quantum criticality. Finally, the comparison with quantum-critical systems without a Fermi surface, such as Dirac fermions at Gross–Neveu transitions discussed in Ref.~\cite{Stangier2026b}, may help clarify how different infrared geometries emerge from distinct classes of strongly interacting many-body systems.

Overall, the derivation presented here highlights that holographic descriptions of quantum matter need not be purely phenomenological. Instead, they can arise as controlled reformulations of microscopic models, offering a geometric perspective on quantum-critical dynamics and pairing in strongly correlated metals.

\acknowledgements
We are grateful to A. V. Chubukov, I. Esterlis, B. Gout{\'e}reaux, S. A. Hartnoll, G.-A.
Inkov, S. Sachdev, K. Schalm, and D. Valentinis for useful discussions.
This work was supported by the German Research Foundation TRR 288-422213477 ELASTO-Q-MAT,
B01 (V.C.S. and J.S.) and grant  SFI-MPS-
NFS-00006741-05 from the Simons Foundation (J.S.).  

\bibliography{refs}

\appendix 
\begin{widetext}
\section{Derivation of the pairing response}
\label{app:A}
In this appendix we analyze the pairing response and hence determine the inverse susceptibility given in Eq.~\eqref{eq:chi_pp}. We will pay particular attention to analyze the mentioned cancellation of the power-law dominant momentum dependent terms.
We analyze:
\begin{equation}
\chi_{\boldsymbol{k}l,l'}\left(\omega,\epsilon\right)=\int_{\boldsymbol{p}}\eta_{l}^{*}\left(\hat{\boldsymbol{p}}\right)\eta_{l'}\left(\hat{\boldsymbol{p}}\right)G_{\boldsymbol{-}\boldsymbol{p}+\frac{\boldsymbol{k}}{2}}\left(-\epsilon+\frac{\omega}{2}\right)G_{\boldsymbol{p}+\frac{\boldsymbol{k}}{2}}\left(\epsilon+\frac{\omega}{2}\right).
\end{equation}
Hence, we analyze the pairing response as function of the center of
gravity momentum $\boldsymbol{k}$, the center of gravity frequency
$\omega$ and the frequency $\epsilon$ that corresponds to the Fourier
transform of the relative time. The fermionic self energy is momentum
independent and given by $\Sigma\left(\epsilon\right)=-i\lambda{\rm sign}\left(\epsilon\right)\left|\epsilon\right|^{1-\gamma}$
such that 
\begin{equation}
G_{\boldsymbol{p}}\left(\epsilon\right)=\frac{1}{i\epsilon-\varepsilon_{\boldsymbol{p}}-\Sigma\left(\epsilon\right)}.
\end{equation}
Then follows
\begin{equation}
\chi_{\boldsymbol{k}l,l'}\left(\omega,\epsilon\right)=\int_{\theta}\int d\varepsilon_{\boldsymbol{p}}\frac{\eta_{l}^{*}\left(\theta\right)\eta_{l'}\left(\theta\right)}{\left(\varepsilon_{\boldsymbol{p}+\frac{\boldsymbol{k}}{2}}+\Sigma\left(\frac{\omega}{2}+\epsilon\right)\right)\left(\varepsilon_{\boldsymbol{p}-\frac{\boldsymbol{k}}{2}}+\Sigma\left(\frac{\omega}{2}-\epsilon\right)\right)}.
\end{equation}
where $\int_{\theta}\cdots=k_{F}\int\frac{d\theta_{\boldsymbol{p}}}{2\pi v_{\boldsymbol{p}}}$
with $v_{\boldsymbol{p}}=\left|\boldsymbol{v}_{\boldsymbol{p}}\right|$
is the magnitude of the velocity $\boldsymbol{v}_{\boldsymbol{p}}$,
which we assume to depend only on the angle $\theta_{\boldsymbol{p}}$.
$l$ and $l'$ are the two quantum numbers and the complete set of
functions is chosen to be orthonormal w.r.t. the scalar product 
\begin{equation}
\left\langle l\mid l'\right\rangle =\int_{\theta}\eta_{l}^{*}\left(\theta\right)\eta_{l'}\left(\theta\right)=\delta_{l,l'}.
\end{equation}

We start with the limit\textbf{ $\boldsymbol{k}=\boldsymbol{0}$}
and $\omega=0$. Then follows
\begin{eqnarray}
\chi_{\boldsymbol{0}l,l}\left(0,\epsilon\right) & = & \delta_{ll'}\int\frac{d\varepsilon}{\left(\varepsilon+\Sigma\left(\epsilon\right)\right)\left(\varepsilon-\Sigma\left(\epsilon\right)\right)}=
  \frac{\pi\delta_{ll'}}{\lambda\left|\epsilon\right|^{1-\gamma}},
\end{eqnarray}
At finite $\omega$ follows instead
\begin{eqnarray}
\chi_{\boldsymbol{0}l,l'}\left(\omega,\epsilon\right) & = & \delta_{ll'}\int\frac{d\varepsilon}{\left(\varepsilon+\Sigma\left(\frac{\omega}{2}+\epsilon\right)\right)\left(\varepsilon+\Sigma\left(\frac{\omega}{2}-\epsilon\right)\right)}=\frac{2\pi\delta_{ll'}}{\lambda}\frac{\theta\left(\left|\epsilon\right|-\frac{\left|\omega\right|}{2}\right)}{\left|\epsilon+\frac{\omega}{2}\right|^{1-\gamma}+\left|\epsilon-\frac{\omega}{2}\right|^{1-\gamma}}.
\end{eqnarray}
If we expand at small $\omega$ it follows
\begin{eqnarray}
\chi_{\boldsymbol{0}l,l'}\left(\omega,\epsilon\right) & = & \frac{\pi\delta_{ll'}}{\lambda\left|\epsilon\right|^{1-\gamma}}\left(1+\frac{1}{8}\gamma\left(1-\gamma\right)\left(\frac{\omega}{\epsilon}\right)^{2}+\cdots\right).
\end{eqnarray}
Next we analyze the limit of small but finite $\boldsymbol{k}$. To
leading order it is sufficient to do this at $\omega=0$.
\begin{eqnarray}
\chi_{\boldsymbol{k}l,l'}\left(0,\epsilon\right) & = & \int\frac{d\theta_{\boldsymbol{p}}}{2\pi v_{\boldsymbol{p}}}\int d\varepsilon_{\boldsymbol{p}}\frac{\eta_{l}^{*}\eta_{l'}}{\left(\varepsilon_{\boldsymbol{p}}+\boldsymbol{v}_{\boldsymbol{p}}\cdot\boldsymbol{k}+\Sigma\left(\epsilon\right)\right)\left(\varepsilon_{\boldsymbol{p}}-\Sigma\left(\epsilon\right)\right)}\\
 & \approx & \chi_{\boldsymbol{0}l,l'}\left(0,\epsilon\right)-\int\frac{d\theta}{2\pi v}\boldsymbol{v}\cdot\boldsymbol{k}\eta_{l}^{*}\eta_{l'}\int\frac{d\varepsilon_{\boldsymbol{p}}}{\left(\varepsilon_{\boldsymbol{p}}+\Sigma\left(\epsilon\right)\right)^{2}}\frac{1}{\varepsilon_{\boldsymbol{p}}-\Sigma\left(\epsilon\right)}\\
 & + & \int\frac{d\theta}{2\pi v}\left(\boldsymbol{v}\cdot\boldsymbol{k}\right)^{2}\eta_{l}^{*}\eta_{l'}\int\frac{d\varepsilon_{\boldsymbol{p}}}{\left(\varepsilon_{\boldsymbol{p}}+\Sigma\left(\epsilon\right)\right)^{3}\left(\varepsilon_{\boldsymbol{p}}-\Sigma\left(\epsilon\right)\right)}
\end{eqnarray}
The energy integrals are given by
\begin{eqnarray}
\int\frac{d\varepsilon_{\boldsymbol{p}}}{\left(\varepsilon_{\boldsymbol{p}}+\Sigma\left(\epsilon\right)\right)^{2}}\frac{1}{\varepsilon_{\boldsymbol{p}}-\Sigma\left(\epsilon\right)} & = & i{\rm sign}\left(\epsilon\right)\frac{\pi}{2\lambda^{2}\left|\epsilon\right|^{2-2\gamma}},\nonumber \\
\int\frac{d\varepsilon_{\boldsymbol{p}}}{\left(\varepsilon_{\boldsymbol{p}}+\Sigma\left(\epsilon\right)\right)^{3}\left(\varepsilon_{\boldsymbol{p}}-\Sigma\left(\epsilon\right)\right)} & = & -\frac{\pi}{4\lambda^{3}\left|\epsilon\right|^{3-3\gamma}}.
\end{eqnarray}
Hence, it follows 
\begin{eqnarray}
\chi_{\boldsymbol{k}l,l'}\left(\omega,\epsilon\right) & = & \frac{\pi\left[\delta_{ll'}\left(1+\frac{\gamma\left(1-\gamma\right)}{8}\left(\frac{\omega}{\epsilon}\right)^{2}\right)-\frac{i{\rm sign}\left(\epsilon\right)\Gamma_{ll'}^{\left(1\right)}}{2\lambda\left|\epsilon\right|^{1-\gamma}}-\frac{\Gamma_{ll'}^{\left(2\right)}}{4\lambda^{2}\left|\epsilon\right|^{2-2\gamma}}\right]}{\lambda\left|\epsilon\right|^{1-\gamma}},
\end{eqnarray}
where $\Gamma_{ll'}^{\left(n\right)}=v_{F}\int_{0}^{2\pi}\frac{d\theta\eta_{l}^{*}\eta_{l'}}{2\pi v\left(\theta\right)}\left(\boldsymbol{v}\cdot\boldsymbol{k}\right)^{n}.$ It holds by definition that $\Gamma_{ll'}^{\left(0\right)}=\delta_{l,l'}$. 

We need the inverse element for the dominant pairing channel to leading
order in $\omega$ and $\boldsymbol{k}$. To this end we write
\begin{equation}
\chi=\chi_{0}+\delta\chi
\end{equation}
Such that 
\begin{eqnarray}
\chi^{-1}  \approx  \chi_{0}^{-1}-\chi_{0}^{-1}\delta\chi\chi_{0}^{-1}+\chi_{0}^{-1}\delta\chi\chi_{0}^{-1}\delta\chi\chi_{0}^{-1}.
\end{eqnarray}

\subsection{Isotropic Fermi surface}

In case of an isotropic Fermi surface we use $\eta_{l}=\frac{1}{\sqrt{2\pi}}e^{il\theta}$, which yields
\begin{eqnarray}
\Gamma_{ll'}^{\left(0\right)} & = & \delta_{ll'}\nonumber \\
\Gamma_{ll'}^{\left(1\right)} & = & \frac{vF}{2}\delta_{ll'+1}\left(k_{x}+ik_{y}\right)+\delta_{ll'-1}\left(k_{x}-ik_{y}\right)\nonumber \\
\Gamma_{ll'}^{\left(2\right)} & = & \frac{v_{F}^{2}}{2}\delta_{ll'}k^{2}+\frac{v_{F}^{2}}{4}\left(\delta_{ll'+2}\left(k_{x}+ik_{y}\right)^{2}+\delta_{ll'-2}\left(k_{x}-ik_{y}\right)^{2}\right).
\end{eqnarray}
We introduce $ K_{x,y}  =  \frac{v_{F}k_{x,y}}{\lambda\left|\epsilon\right|^{1-\gamma}}$ and $\chi_{0}  =  \frac{\pi}{\lambda\left|\epsilon\right|^{1-\gamma}} $ such that 
\begin{eqnarray}
\chi_{\boldsymbol{k}l,l'}\left(\omega,\epsilon\right) & = & \chi_{0}\left(\delta_{ll'}\left(1+\frac{\gamma\left(1-\gamma\right)}{8}\left(\frac{\omega}{\epsilon}\right)^{2}-\frac{K^{2}}{8}\right)-\frac{i{\rm sign}\left(\epsilon\right)\delta_{ll'\pm1}\left(K_{x}\pm iK_{y}\right)}{4}\right.
 - \left.\frac{\delta_{ll'\pm2}\left(K_{x}\pm iK_{y}\right)^{2}}{16}\right)
\end{eqnarray}
It follows for the diagonal elements of the inverse
\begin{equation}
\left.\chi_{\boldsymbol{k}}^{-1}\left(\omega,\epsilon\right)\right|_{ll}=\chi_{0}^{-1}\left(1-\frac{\gamma\left(1-\gamma\right)}{8}\left(\frac{\omega}{\epsilon}\right)^{2}+\frac{K^{2}}{8}\right)-\chi_{0}^{-1}\frac{K^{2}}{8}
\end{equation}
Hence, for an isotropic Fermi surface follows that the leading gradient
terms cancel exactly. Next we show that this result also holds for  a generic Fermi surface geometry. 

\subsection{Generic Fermi surface}
Now we analyze 
\begin{equation}
\Gamma_{ll'}^{\left(n\right)}\left(\boldsymbol{k}\right)=v_{F}\int_{0}^{2\pi}\frac{d\theta\eta_{l}^{*}\eta_{l'}}{2\pi v\left(\theta\right)}\left(\boldsymbol{v}\cdot\boldsymbol{k}\right)^{n}.
\end{equation}
We consider a system with inversion symmetry, which for $d=2$ really means $C_{2z}$ rotation invariance. Then follows  that the eigen-functions $\eta_{l}$ have well defined parity. For even $n$ holds that only the same parity functions contribute.
For odd $n$ follows instead that $\eta_{l}$ and $\eta_{l'}$ must have opposite
parity. Like in our above analysis for spherical Fermi surfaces follows
that we  only need to determine the diagonal elements
$\Gamma_{ll}^{\left(2\right)}\left(\boldsymbol{k}\right)$ and off-diagonal
elements of $\Gamma_{ll}^{\left(1\right)}\left(\boldsymbol{k}\right)$. 

We first consider
\begin{eqnarray}
\Gamma_{ll}^{\left(2\right)}\left(\boldsymbol{k}\right) & = & v_{F}\int_{0}^{2\pi}\frac{d\theta\left|\eta_{l}\right|^{2}}{2\pi v\left(\theta\right)}\left(v_{x}k_{x}+v_{y}k_{y}\right)^{2}=v_{F}\sum_{\alpha\beta}k_{\alpha}k_{\beta}\int_{0}^{2\pi}\frac{d\theta\left|\eta_{l}\right|^{2}}{2\pi v\left(\theta\right)}v_{\alpha}v_{\beta}\nonumber \\
 & = & \sum_{\alpha\beta}k_{\alpha}k_{\beta}\left\langle l\left|v_{\alpha}v_{\beta}\right|l\right\rangle _{FS}.
\end{eqnarray}
For example, in a tetragonal system holds
\begin{eqnarray}
\left\langle l\left|v_{x}^{2}\right|l\right\rangle _{FS} & = & v_{F}\int_{0}^{2\pi}\frac{d\theta\left|\eta_{l}\right|^{2}}{2\pi v\left(\theta\right)}v_{x}^{2}=v_{F}\int_{0}^{2\pi}\frac{d\theta\left|\eta_{l}\right|^{2}}{2\pi v\left(\theta\right)}v_{y}^{2}=\left\langle l\left|v_{y}^{2}\right|l\right\rangle _{FS}\nonumber \\
\left\langle l\left|v_{x}v_{y}\right|l\right\rangle _{FS} & = & 0.
\end{eqnarray}
To leading order, the off-diagonal elements are governed by the term
\begin{equation}
\Gamma_{ll'}^{\left(1\right)}\left(\boldsymbol{k}\right)=v_{F}\int_{0}^{2\pi}\frac{d\theta\eta_{l}^{*}\eta_{l'}}{2\pi v\left(\theta\right)}\left(v_{x}k_{x}+v_{y}k_{y}\right)=\sum_{\alpha}\left\langle l\left|v_{\alpha}\right|l'\right\rangle _{FS}k_{\alpha}.
\end{equation}
It then follows:
\begin{eqnarray*}
\chi_{\boldsymbol{k}l,l'}\left(\omega,\epsilon\right) & = & \chi_{0}\delta_{ll'}\left(1+\frac{\gamma\left(1-\gamma\right)}{8}\left(\frac{\omega}{\epsilon}\right)^{2}-\frac{1}{4}\sum_{\alpha\beta}K_{\alpha}K_{\beta}\left\langle l\left|v_{\alpha}v_{\beta}\right|l\right\rangle _{FS}\right),\\
 & - & i\frac{\chi_{0}}{2}{\rm sign}\left(\epsilon\right)\sum_{\alpha}K_{\alpha}\left\langle l\left|v_{\alpha}\right|l'\right\rangle _{FS}
\end{eqnarray*}
If we now analyze the diagonal elements of the inverse, it follows
\begin{eqnarray}
\chi_{\boldsymbol{k}}^{-1}\left(\omega,\epsilon\right)_{l,l} & = & \chi_{0}^{-1}\left(1-\frac{\gamma\left(1-\gamma\right)}{8}\left(\frac{\omega}{\epsilon}\right)^{2}+\frac{1}{4}\sum_{\alpha\beta}K_{\alpha}K_{\beta}\left\langle l\left|v_{\alpha}v_{\beta}\right|l\right\rangle _{FS}\right),\nonumber \\
 & - & \frac{1}{4}\chi_{0}^{-1}\sum_{l'\alpha\beta}\left\langle l\left|v_{\alpha}\right|l'\right\rangle _{FS}\left\langle l'\left|v_{\beta}\right|l\right\rangle _{FS}K_{\alpha}K_{\beta}
\end{eqnarray}
The terms proportional to $K_{\alpha}K_{\beta}$ cancel again.
Including terms that go beyond the linear expansion of the dispersion then yields the formally sub-leading terms $\sim \boldsymbol{k}^2/k^2_{\rm F}$ given in Eq.~\eqref{eq:chi_pp}.
\end{widetext}

\end{document}